\documentclass[twocolumn,dvipsnames]{aastex62}
\pdfoutput=1 
\usepackage{amsmath,amstext}
\usepackage{apjfonts}
\usepackage[figure,figure*]{hypcap}
\usepackage{ulem}
\usepackage{xspace}
\usepackage[utf8]{inputenc}
\usepackage[T1]{fontenc}
\usepackage{graphicx}
\usepackage{natbib}
\usepackage{amssymb}
\usepackage{amsmath}
\usepackage{array,booktabs}
\usepackage{mathptmx}
\usepackage{booktabs}
\usepackage{hyperref}
\usepackage{float}
\newcommand{\msun}{\,{M_{\odot}}}

\newcommand{\erg}{\,{\rm erg}}

\newcommand{\vrotz}{\,{\omega_0}}

\newcommand{\appropto}{\mathrel{\vcenter{
			\offinterlineskip\halign{\hfil$##$\cr
				\propto\cr\noalign{\kern2pt}\sim\cr\noalign{\kern-2pt}}}}}
\usepackage{xcolor}
\usepackage[normalem]{ulem} 
\newcommand{\myvec}[1]{\textbf{\textrm{#1}}}

\shorttitle{Wobbling and hybrid composition lGRB jets}
\shortauthors{Gottlieb et al.}

\begin{document}
	\title{Black hole to photosphere: 3D GRMHD simulations of collapsars reveal wobbling and hybrid composition jets}
\author[0000-0003-3115-2456]{Ore Gottlieb}
\email{ore@northwestern.edu}
\affiliation{Center for Interdisciplinary Exploration \& Research in Astrophysics (CIERA), Physics \& Astronomy, Northwestern University, Evanston, IL 60202, USA}

\author{Matthew Liska}
\affiliation{Institute for Theory and Computation, Harvard University, 60 Garden Street, Cambridge, MA 02138, USA; John Harvard Distinguished Science and ITC}

\author[0000-0002-9182-2047]{Alexander Tchekhovskoy}
\affiliation{Center for Interdisciplinary Exploration \& Research in Astrophysics (CIERA), Physics \& Astronomy, Northwestern University, Evanston, IL 60202, USA}

\author{Omer Bromberg}
\affiliation{School of Physics and Astronomy, Tel Aviv University, Tel Aviv 69978, Israel}

\author[0000-0002-6883-6520]{Aretaios Lalakos}
\affiliation{Center for Interdisciplinary Exploration \& Research in Astrophysics (CIERA), Physics \& Astronomy, Northwestern University, Evanston, IL 60202, USA}

\author[0000-0003-1503-2446]{Dimitrios Giannios}
\affiliation{Department of Physics and Astronomy, Purdue University, West Lafayette, IN 47907, USA}

\author[0000-0002-9371-1447]{Philipp M\"{o}sta}
\affiliation{GRAPPA, Anton Pannekoek Institute for Astronomy and Institute of High-Energy Physics, University of Amsterdam, Science Park 904, 1098 XH Amsterdam, The Netherlands}

	\begin{abstract}
	
		Long-duration $\gamma$-ray bursts (GRBs) accompany the collapse of massive stars and carry information about the central engine. However, no 3D models have been able to follow these jets from their birth by a black-hole (BH) to the photosphere. We present the first such 3D general-relativity magnetohydrodynamic simulations, which span over 6 orders of magnitude in space and time. The collapsing stellar envelope forms an accretion disk, which drags inwardly the magnetic flux that accumulates around the BH, becomes dynamically important and launches bipolar jets. The jets reach the photosphere at $\sim10^{12}$ cm with an opening angle $\theta_j\sim6^\circ$ and a Lorentz factor $\Gamma_j\lesssim30$, unbinding $\gtrsim90\%$ of the star. We find that (i) the disk-jet system spontaneously develops misalignment relative to the BH rotational axis. As a result, the jet wobbles with an angle $\theta_t\sim12^\circ$, which can naturally explain quiescent times in GRB lightcurves. The effective opening angle for detection $\theta_j+\theta_t$ suggests that the intrinsic GRB rate is lower by an order of magnitude than standard estimates. This suggests that successful GRBs may be rarer than currently thought and emerge in only $\sim0.1\%$ of supernovae Ib/c, implying that jets are either not launched or choked inside most supernova Ib/c progenitors. (ii) The magnetic energy in the jet decreases due to mixing with the star, resulting in jets with a hybrid composition of magnetic and thermal components at the photosphere, where $\sim 10\%$ of the gas maintains magnetization $\sigma\gtrsim0.1$. This indicates that both a photospheric component and reconnection may play a role in the prompt emission.
		
	\end{abstract}
	
	\section{Introduction}\label{sec:introduction}
	
	Long-duration gamma-ray bursts (lGRBs) are thought to emerge when collimated relativistic jets escape a collapsing massive star  \citep[collapsar;][]{Woosley1993,Macfadyen1999}. The prospects of learning from this phenomenon about the properties of GRB progenitor stars and the physics of their central engines inspired analytic studies of the propagation of relativistic jets in stars \citep[e.g.,][]{Matzner2003,Lazzati2005,Bromberg2011}. The nonlinear behavior of jets implies that numerical tools are essential for modeling the jet evolution before and after it breaks out from the star. This realization motivated the first 2D hydrodynamical simulations of jet propagation in stars \citep[e.g.][]{Aloy2000,MacFadyen2001,Zhang2003,Zhang2004,Mizuta2006,Morsony2007,Morsony2010,Bucciantini2009,Komissarov2009,Lazzati2009,Mizuta2009,Nagakura2011,Mizuta2013,Duffell2015}. Despite achieving a significant progress in understanding lGRBs, these models suffer from several major drawbacks:
	
	(i) Axisymmetry: 2D models artificially impose axial symmetry, which suppresses nonaxisymmetric modes and leads to numerical artifacts. The growth in computational power and improvements in algorithms has enabled 3D modeling of hydrodynamic jets in collapsars \citep{Lopez-camara2013,Lopez-camara2016,Ito2015,Ito2019,Harrison2018,Gottlieb2019,Gottlieb2020a,Gottlieb2021a}. These studies demonstrated that 3D jets can feature utterly different behavior than 2D, motivating full 3D jet studies.
	
	(ii) Magnetization: Relativistic jet launching from stellar engines can either be driven electromagnetically by the rotation of a compact object -- a black hole \citep[BH;][]{Blandford1977} or a neutron star \citep[][]{Goldreich1969, Usov1992,Thompson1994,Metzger2011} -- or thermodynamically by the pair plasma produced by annihilation of neutrinos originating in the accretion disk \citep[e.g.][]{Eichler1989,Paczynski1990,Macfadyen1999}. It is becoming increasingly clear that the latter energy source falls short of the enormous amounts of energy required to power lGRB jets, lending support to the idea that the outflow is magnetically powered  \citep[e.g.,][]{Kawanaka2013,Leng2014,Liu2015}. Numerical simulations of magnetized jets are challenging, and only a few 3D models of magnetized jet evolution have been performed over the years \citep[e.g.,][]{Mignone2010,Porth2013,Bromberg2016,Striani2016}. Simulations of magnetic jets propagating to the stellar surface were performed only recently, although with some limitations: Either the jet was injected with subdominant magnetic energy \citep[e.g.][]{Gottlieb2020b,Gottlieb2021b} or the central engine duration was too short to allow a successful breakout of relativistic material from the star \citep[][hereafter G22]{Gottlieb2022a}. In both cases, the relativistic outflows were not modeled beyond the breakout phase. Nevertheless, these works highlighted the importance of jet magnetization in jet stability, structure, and propagation.
	
	(iii) All aforementioned studies of jets in stars (apart from G22 and the 2D simulations of \citealt{Komissarov2009}) prescribed the jet launching from a grid boundary rather than including self-consistent launching via the rotation of a magnetized central compact object. Because each chapter in the jet journey influences the following ones, connecting the underlying physics at the central BH with observations can only be achieved through a complete modeling of the entire jet evolution, from a self-consistent jet launching by a spinning central BH to the emission zone.
	
	G22 performed 3D general-relativity magnetohydrodynamic (GRMHD) simulations of collapsars to study the effect of the progenitor structure on the jet launching and breakout. They found that the type of outflow depends on the initial magnetization, rotation, and mass density profiles of the star. If the magnetic field is too weak, a relativistic jet is not launched, and instead a standing accretion shock dominates the outflow. If the stellar rotation is too slow, a quasi-spherical outflow is driven by the engine, rather than a collimated jet. If both the magnetic field is strong and the rotation is fast enough to allow for an accretion disk to form, a relativistic jet is launched. Their simulations showed that the jet is subject to strong magnetic dissipation at the collimation nozzle, which renders the jet essentially hydrodynamic upon breakout from the star. However, the spatial resolution at the narrow collimation nozzle was marginal to verify the robustness of this result. Additionally, it showed that some of the shocked gas remains bound and free-falls toward the BH before it is deflected by the jet to fall onto the accretion disk. This tilts the accretion disk and, subsequently, the jet orientation.
	
	Here, we build on the results of G22 to present a long-awaited 3D GRMHD simulation that follows the jets for their entire evolution: from a self-consistent launching near a BH to an unprecedented distance of $\sim 10^6$ gravitational radii. To the best of our knowledge, this is the first simulation that features 3D relativistically magnetized jets that reach relativistic Lorentz factors after breakout from the star. With the highest-resolution GRB simulation to date, we investigate the magnetic dissipation at the collimation nozzle and study the observational implications of the disk-jet tilt on the post-breakout outflow at $\gtrsim 10$ stellar radii.
	
	The structure of the paper is as follows. In \S\ref{sec:setup} we recap the numerical setup of G22 and discuss the modifications required to produce steady relativistic jets that operate for the entire duration of the simulation, $\sim 18$ s. In \S\ref{sec:evolution} we outline the main physical processes that take place at the BH horizon, the magnetic dissipation, the jet tilt and post-breakout structure. In \S\ref{sec:emission} we discuss the implications of our results for GRB rates, the variability and quiescent times in GRB lightcurves, and the powering mechanism for the GRB prompt emission. In \S\ref{sec:conclusions} we summarize the results.
	
	\section{Setup}\label{sec:setup}
	
	G22 performed 3D GRMHD simulations of self-consistent jet launching in collapsars, using the 3D GPU-accelerated code \textsc{h-amr} \citep{Liska2019}. They modeled the conditions for jet launching in collapsars, and connected the progenitor magnetic field, rotation and mass density profiles to the emerging jet properties.
	However, the central engine in G22 did not launch jets for a sufficiently long time to match the observed durations of lGRBs ($\sim2{-}100$~s) and allow for the jets to break out of the star. G22 suggested that a change in the stellar magnetic field configuration can increase the duration of the active jet phase.
	Here we use nearly identical initial conditions to G22, but modify the initial magnetic field configuration to attain a longer active jet phase.
	
	The initial conditions of the simulation adopt a Kerr BH of mass $ M_{\rm BH,0} = 4M_\odot $ and dimensionless spin $ a = 0.8 $, embedded in a cold, compact Wolf-Rayet--like star of radius $ R_\star = 4\times 10^{10} $ cm and mass $ M_\star \approx 14\msun $. The density profile of the progenitor star at the beginning of the simulation (upon BH formation) is
	\begin{equation}\label{eq:progenitor}
		\rho(r) = \rho_0 \left(\frac{r}{r_g}\right)^{-\alpha}\left(1-\frac{r}{R_\star}\right)^3,
	\end{equation}
	where $ \rho_0 $ is set by the requirement that $ M_\star =\int_0^{R_*}\rho(r)dV$ and $ r_g $ is the BH gravitational radius.
	G22 found that moderate values of $ \alpha \sim 1 $ can be consistent with many GRB observables: signal duration of $ \gtrsim 10 $ s, jet luminosity $ L_j \approx 10^{51}~{\rm erg~s^{-1}} $, and the absence of long-term trends in the time evolution of the lightcurve. We adopt $ \alpha = 1.5 $ that, as we show here, satisfies all of these observables.
	
	The specific angular momentum profile of the stellar envelope is spherically symmetric, increasing to $ 70r_g $, and then plateaus:
	\begin{equation}
		l(r) =
		\begin{cases}
			\vrotz\left(\frac{r^2}{r_g}\right)^2 & r < 70r_g \\
			&\\
			\vrotz (70^2r_g)^2 & r > 70r_g
		\end{cases} ~,
	\end{equation}
	where $\vrotz$ is constant.
	We adopt a uniform vertical magnetic field $\myvec{B}_{_{\rm core}}=B_0\hat{z}$ inside the stellar core of radius $ r_c = 10^8 $ cm. While G22 used a magnetic dipole outside the magnetic core, here we employ a shallower magnetic profile with a vector potential. The reason for this choice lies in the need to advect enough magnetic flux to the BH also at late times, so that the central engine remains active throughout the entire duration of the simulation. G22 showed that the jet duration scales with the ratio of the fastest growing magnetorotational instability (MRI) wavelength mode to the disk scale height $ \propto B^{1/2} $, where $ B $ is the magnetic field magnitude. Thus, a slow radial decay in the magnetic field allows a longer jet launching duration. Overall, the magnetic vector potential is
	\begin{equation}\label{eq:Bprofile}
		\myvec{A}=A_{\hat\phi}(r,\theta)\hat{\phi} = \mu\frac{{\rm sin}\theta}{r}\cdot {\rm max}\bigg[\frac{r^2}{r^2 + r_c^2} - \left(\frac{r}{R_\star}\right)^3,0\bigg]\hat{\phi}~,
	\end{equation}
	where $ \mu \approx B_0r_c^2 $ is the magnetic moment of the uniformly magnetic core, and the second term determines how fast the magnetic field drops near the stellar edge. The choice of $ B_0 $ is such that the maximum magnetization $ \sigma \equiv b^2/(4\pi\rho c^2)$ in the star is $\max\sigma\sim 10^{-1.5} $; here $b$ is the comoving magnetic field strength. These values correspond to maximal magnetic field in the core of $ B \sim 10^{12.5} $ G, which is required for the jet to overcome the ram pressure of the infalling gas, as found by G22. This choice of initially strong magnetic field implicitly assumes that the magnetic field was amplified prior to the BH formation (onset of the simulation), presumably following the rapid free-falling plasma that accumulates in the core after the initial collapse.
	
	The simulation is performed with an ideal equation of state with adiabatic index of $ 4/3 $, appropriate for relativistic or radiation dominated gas. For numerical stability purposes we set density floors in the code by setting the maximal magnetization in the simulation to $ \sigma_0 = 15 $. The maximum magnetization is roughly the maximal asymptotic velocity that a fluid element can reach (up to a correction factor of order unity for the thermal energy of launched hot jets). We note that these values are much lower than those of GRBs. Therefore, we also carry out an identical simulation with $ \sigma_0 = 200 $, with which we address the potential effects $ \sigma_0 $ on our results.
	The plasma magnetization in this case is more sensitive to an artificial injection of floor mass density in regions where $\sigma$ approaches the limit set by $\sigma_0$. Nevertheless, we can use the outcome both as a consistency check for the results of the simulation with  $\sigma_0=15$, and as a lower limit to the expected magnetization outside of the star. Throughout the paper, we compare the results of the two simulations and address the possible interpretations.
	
	For the numerical integration we use a local adaptive time-step and 4 levels of adaptive mesh refinement (AMR). We use a spherical grid with a logarithmic cell distribution in the radial direction and uniform distributions in $ \hat{\theta} $ and $ \hat{\phi} $ directions. The radial grid extends from just inside the event horizon, $ \sim 6\times 10^5 $ cm, to $ \sim 6\times 10^{11} $ cm with numerical resolution at the base AMR level of $N_r\times N_\theta\times N_\phi = 384\times96\times192 $ cells, in the $r$-, $\theta$-, and $\phi$-directions, respectively. We use a novel refinement criterion that at each radius $r$ calculates the jet and cocoon (see its definition below) half-opening angles based on the specific entropy of the fluid, and if either one of them contains less than the desired number of cells, $ \Delta N_\theta = 96 $ or $ \Delta N_\phi = 192 $, the grid refines to the next AMR level, until it reaches the desired number of cells across each dimension, up to 4 levels of refinement. Therefore, the effective number of cells at the maximum level of refinement is $ N_r \times N_\theta \times N_\phi = 6144 \times 1536 \times 3072 \approx 3\times 10^{10} $. To avoid numerical artifacts associated with the jet propagation along the polar axis, we tilt the metric by $ 90^\circ $ such that the angular momentum orbital plane is the $ \hat{y}-\hat{z} $ plane, and the stellar rotation is around the $ \hat{x} $-axis. To avoid confusion, in the text and figures we refer to the $ \hat{z} $-axis and $ \theta $ as the stellar rotation axis and the polar angle relative to the rotation axis, respectively.
	
	\section{Results}\label{sec:evolution}
	
	\subsection{Early jet evolution}\label{sec:prebreakout}
	
	We start our simulation after the collapse of the inner stellar core to a BH, and simulate the subsequent collapse of the stellar envelope onto the central BH. A few milliseconds after the onset of the simulation, an accretion disk forms and powerful bipolar relativistic jets are launched. Our choice of the initial stellar progenitor magnetic field and density profiles ensures that the jet power is sufficient to overcome the ram pressure of the infalling gas and sustain a steady outflow from the BH vicinity. As the jet propagates through the stellar envelope, it forms a double shocked layer cocoon (e.g. Figure~\ref{fig:3djet}). The high-density outer part of the cocoon consists of the stellar material heated up by the forward shock, while the low-density inner cocoon consists of the jet material heated up by the reverse shock \citep{Bromberg2011}. The pressure of the cocoon confines and collimates the jets. This collimation is effective at large radii; at smaller radii the jets are confined by the accretion disk winds.
	
\begin{figure}
		\centering
		\includegraphics[scale=0.22,trim=0 0 0 0]{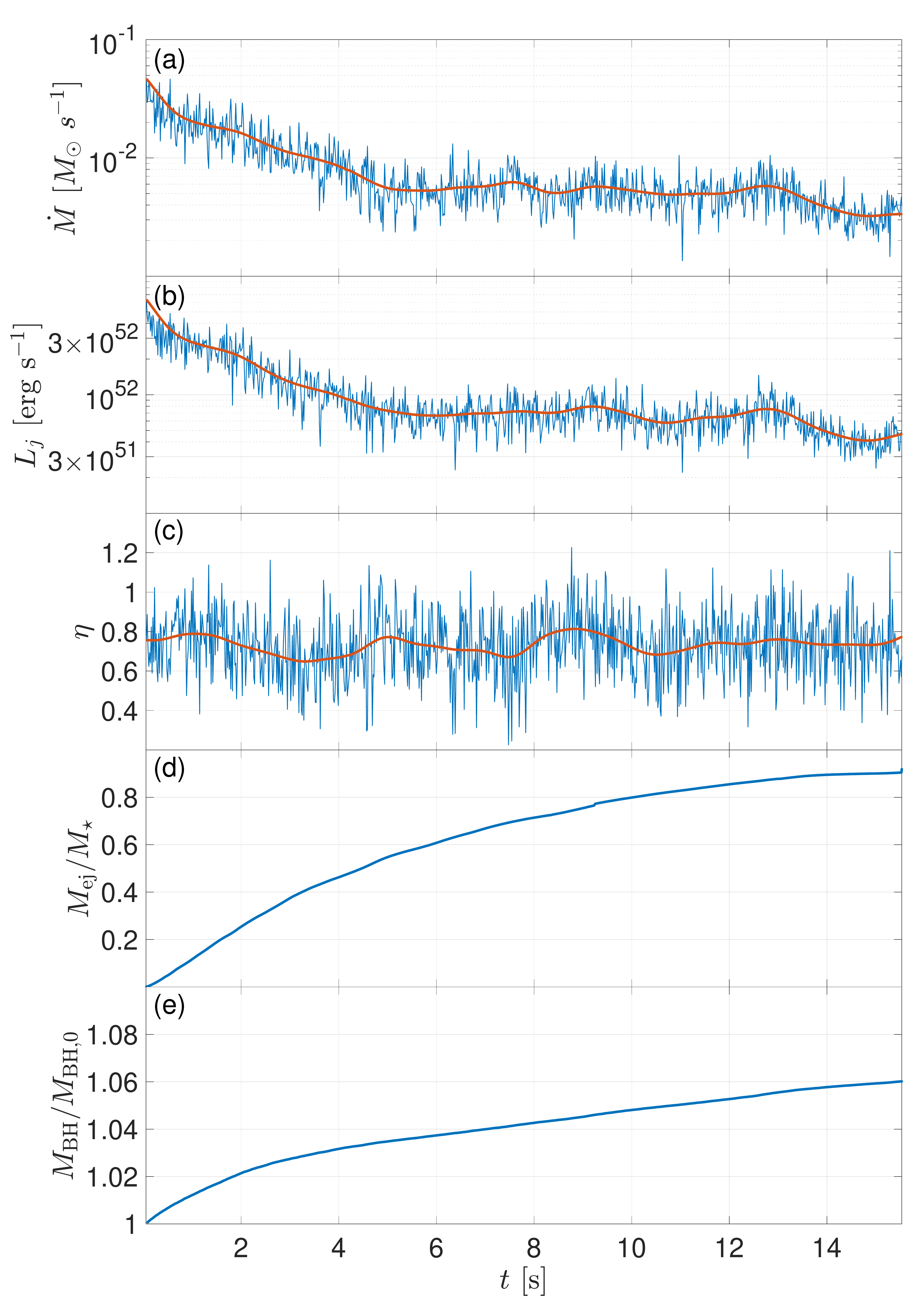}
		\caption[]{Time evolution in the simulation with $ \sigma_0 = 15 $: \textbf{Panel (a):} Accretion onto the BH initially decreases due to the inflation of the cocoon, before plateauing after a few seconds. \textbf{Panel (b):} The jet luminosity, which is calculated as the radial energy flux excluding the rest energy flux, scales proportionally to the mass accretion rate. \textbf{Panel (c):} The jet launching efficiency, $\eta$, remains of order unity at all times, signaling that the system is accreting in the MAD regime. The jet-cocoon outflow unbinds most of the stellar envelope, ejecting $ \sim 90\% $ of the total stellar mass by $ t \approx 16$~s (\textbf{panel (d)}). As most of the stellar mass becomes unbound early on, the available gas for accretion is limited such that the BH mass $ M_{\rm BH} $ increases only by a few percent over the duration of the simulation (\textbf{panel (e)}). The red lines show moving averages of the highly variable quantities.
		}
		\label{fig:properties}
	\end{figure}

    The cocoon regulates the jet power by suppressing the rate at which the stellar envelope feeds the accretion disk. This reduces the accretion rate from the expected scaling of $ \dot{M} \propto t^{1-2\alpha/3} $ (in the case of a free falling stellar envelope with a power law density profile index $ \alpha $) to $ \dot{M} \sim t^{2(1-\alpha)/3} $ (G22). Figure~\ref{fig:properties}(a) shows that the time evolution of the mass accretion rate at $ r = 5r_g $ decays as expected (for our choice of $\alpha = 1.5$), $ \dot{M} \propto t^{-1/3} $ during the first few seconds. The jet breaks out from the star after $ \sim 3 $ s, and when the jet head reaches $ \sim 2 R_\star $ at $ \sim 5 $ s, the jet no longer energizes the cocoon. From this point on, the lateral velocity of the cocoon decreases, and by the time the cocoon shocks the entire star, its velocity has dropped by a factor of the jet opening angle $ \theta_j $ \citep{Eisenberg2022}. The decay in the cocoon expansion velocity moderates its effect on the mass accretion rate, which nearly plateaus after $ \sim 5 $ s.
	
	Figures~\ref{fig:properties}(b) and (c) depict the jet luminosity $ L_j $ at $ r = r_g $, and the jet power efficiency $ \eta = L_j/\dot{M}c^2 $ on the horizon (as defined in G22), respectively. The jet maintains an order unity energy efficiency of $\eta=0.7$ at all times, from the moment it is launched until the end of the simulation, implying that it is accompanied by a magnetically arrested disk \citep[MAD;][]{tch11}. As a result, the jet luminosity on the horizon follows the accretion rate as $ L_j \sim \dot{M}c^2 $, as shown in Fig.~\ref{fig:properties}(b). The jet power exhibits intermittency on short timescales ($10~{\rm ms} \lesssim \Delta t \lesssim 100 $ ms), comparable to those observed in GRB lightcurves, with amplitude variations of about half an order of magnitude.
	
	As the jet-cocoon structure expands, it carries with it an increasing amount of stellar material, which will later escape from the star once the cocoon breaks out. Toward the end of the simulation, $ \sim 90\% $ of the stellar mass is affected by the cocoon and becomes hydrodynamically unbound (Fig.~\ref{fig:properties}d). This suggests that in collapsars, the BH mass at the time of the disk formation $ M_{\rm BH,0} $ is similar to the final BH mass $ M_{\rm BH} $, as can be seen in Fig.~\ref{fig:properties}(e). We note that in stars with a steeper density profile ($ \alpha \gtrsim 2 $), the accretion rate is highest at early times, before the jet unbinds most of the star. It is therefore possible that in such stars the BH accretes a comparable amount of gas to its initial mass (G22), thereby significantly reducing its dimensionless spin and jet launching efficiency \citep{Tchekhovskoy2012}.
	We emphasize that our numerical simulation includes neither self-gravity nor the internal energy of the stellar envelope, and therefore, the amount of bound and unbound mass could be different when the simulation properly accounts for these effects. However, because $ M_\star \sim 3M_{\rm BH,0} $, the free-fall time and escape velocity would only change by up to factors of, respectively, $ \lesssim 1.5$ and $2$ in the outermost layers had self-gravity been included. Moreover, in the outermost layers, where most of the stellar mass lies and self-gravity is important, the shocked gas is accelerated to the highest velocities by the relativistic outflow, and thus, the dynamics in the pre-shocked gas is likely to have a weak effect.
	
	G22 showed that when the bound parts of cocoon material fall toward the BH, it hits the polar outflow, gets deflected, and hits the accretion disk, kicking it out of alignment with the BH. Consequently, the accretion disk tilt changes. If these stochastic changes add up constructively, the disk can develop a substantial tilt values of up to $ \sim 60^\circ $ (see movie at \url{http://www.oregottlieb.com/collapsar.html}). Because the jet is launched perpendicularly to the disk \citep{2018MNRAS.474L..81L}, changes in disk tilt also alter the jet orientation.
	
\begin{figure}
		\centering
		\includegraphics[scale=0.33,trim=0 0 0 0]{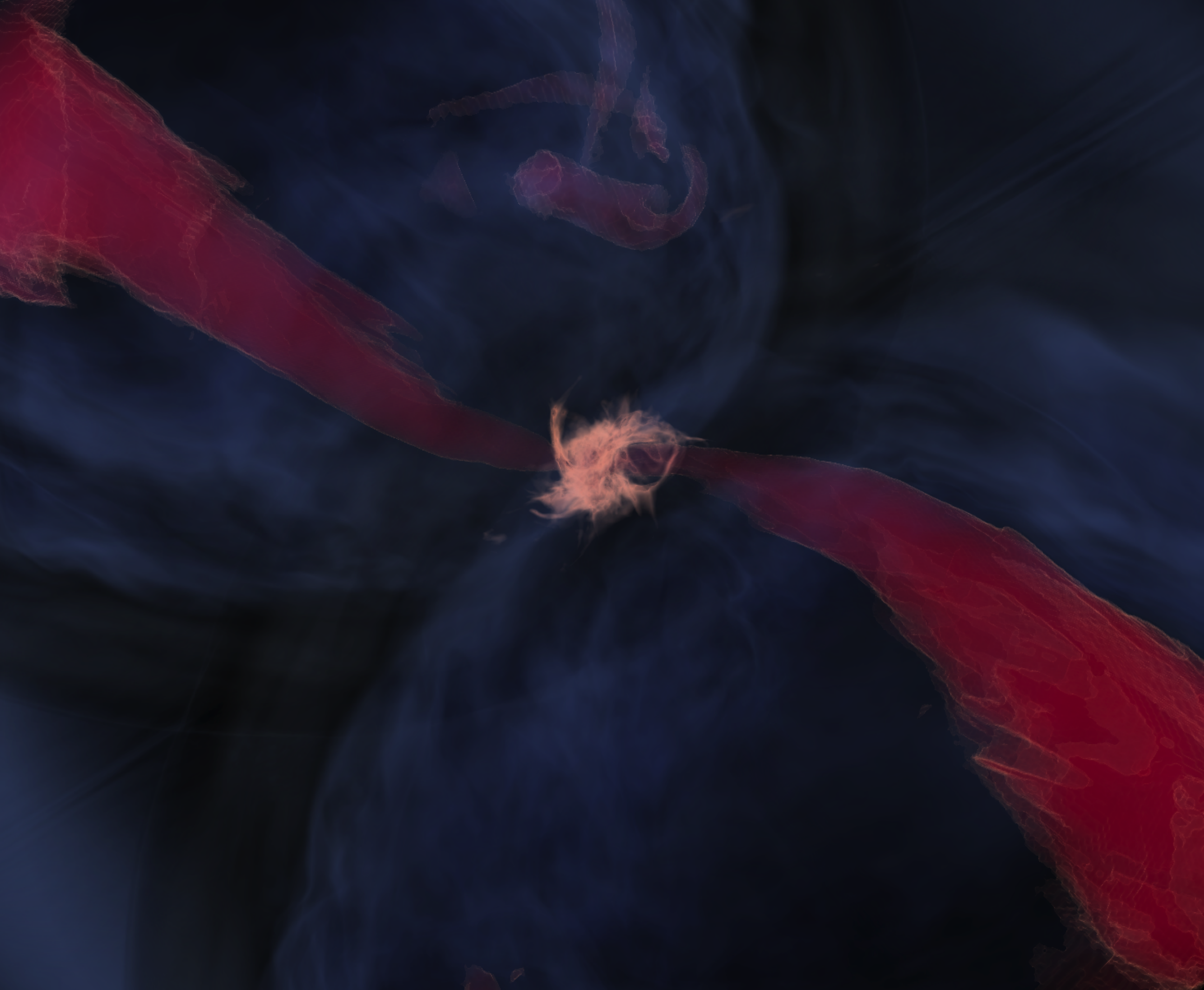}
		\includegraphics[scale=0.344,trim=0 0 0 0]{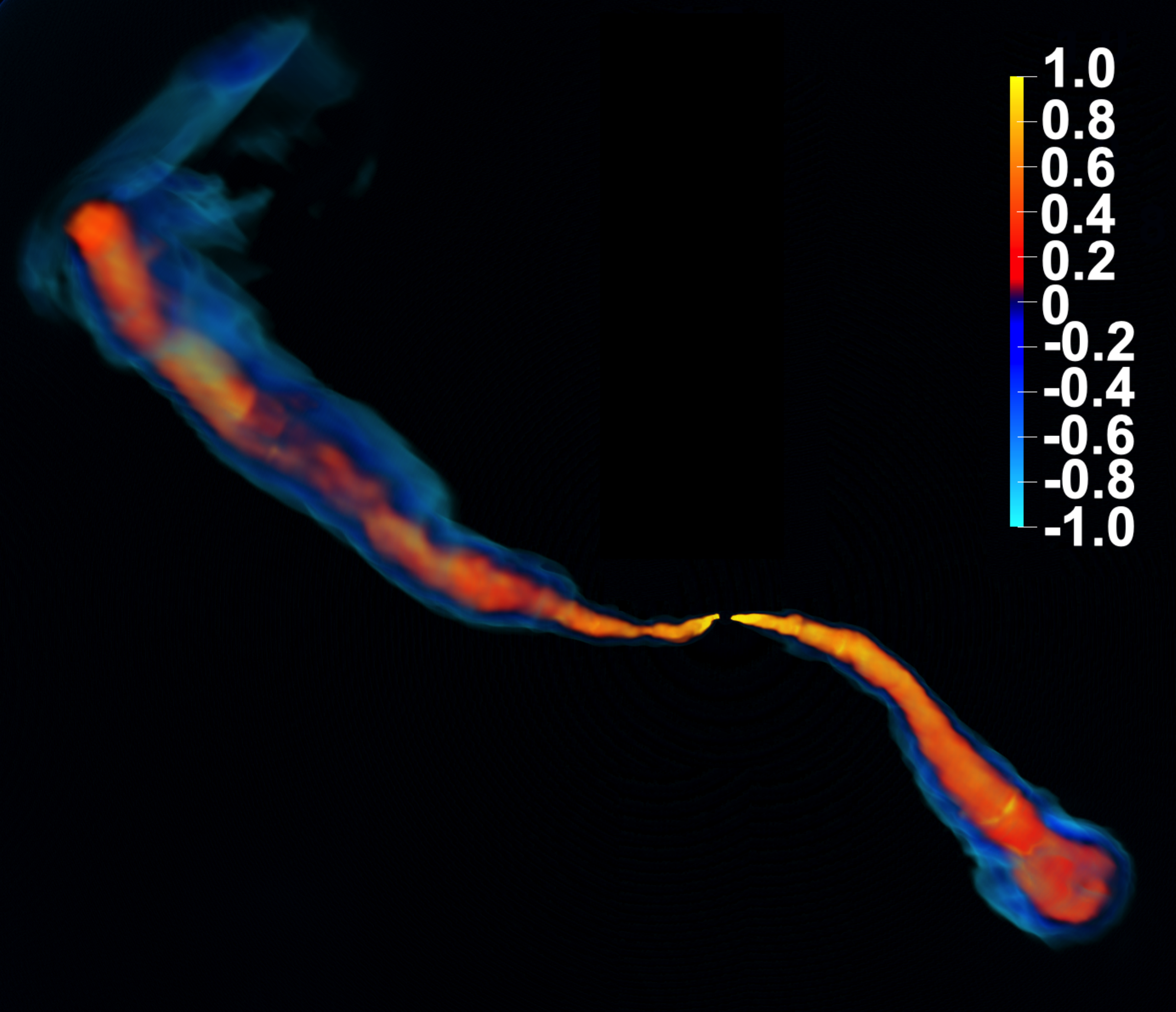}
		\caption[]{
			Top: 3D rendering of the $ \sigma_0 = 15 $ jet in the inner $ 2\times 10^9 $ cm shows a significant tilt of the disk (orange) and jet (red) axis (horizontal) with respect to the rotation axis of the star at $ 45^\circ $. Although the jet is launched at different orientations at different times, it is deflected by the heavy outer cocoon material (blue gray) that engulfs the jet region toward the rotation axis at $ 45^\circ $. Bottom: 3D rendering of the logarithm of jet magnetization shows the deflection of the jet propagation and a drop in the magnetization from $ \sigma \sim 10 $ to $ \sigma \sim 1 $. Here, the jet head is located at $ r = 0.1R_\star $. Movies showing the full evolution of the disk-jet tilt are available at \url{http://www.oregottlieb.com/collapsar.html}.
		}
		\label{fig:early3d}
	\end{figure}

	The top panel of Fig.~\ref{fig:early3d} depicts a 3D rendering of the disk (orange) and jets (red) embedded in the cocoon (gray blue). The jets fly out along the disk rotational axis, which is horizontal in the figure and tilted by $ \sim 45^\circ $ with respect to the stellar rotation axis, which points from the top-left to the bottom-right corner of the panel. One might expect that a jet that is launched in different directions will be soon choked as it needs to drill a new path for each new direction. However, our simulation shows that as soon as the tilted jets are launched, they are deflected toward the low-density regions drilled by the earlier non-tilted jets along the original angular momentum axis of the disk. This behavior is seen in the figure as the tilted jets gradually curve toward an angle of $ \sim 45^\circ $ as soon as they run into the dense edges of the outer cocoon (gray and blue). While the jets' deviation from the axis is moderated with time, they still feature a $ \sim 0.2 $ rad tilt with respect to the polar axis upon breakout (\S\ref{sec:tilt}). The idea of jittering jets in a star was first proposed by \citet{Papish2011} as an efficient way for the jets to explode the entire star. We find that although our simulation features jittering jets, they are soon deflected back toward the low-density regions so that the jets successfully break out from the star. Had the jets maintained the altered orientation with which they are launched, they would have failed to pierce through the star and would have been choked. In \S\ref{sec:tilt} and \S\ref{sec:emission} we study the consequences of disk tilt on the post-breakout stage and its observational implications, respectively.

	\subsection{Magnetic dissipation}\label{sec:dissipation}
	
	3D RMHD simulations of highly magnetized jets in a dense medium discovered that they develop current driven instabilities, primarily the magnetic kink instability, once they run into the dense material and recollimate, forming a magnetic nozzle. This is accompanied by the dissipation of their magnetic energy into heat \citep[][]{Bromberg2016}. G22 recently found that a similar process also takes place in GRMHD simulations of collapsars: The bipolar jets dissipate most of their magnetic energy and become mildly magnetized above the nozzle. However, in G22 the resolution at the collimation nozzle was marginal, with only $ \lesssim 6 $ cells across the jet half-opening angle near the nozzle. Here, our simulation is structured such that at any radius and time, the jet half-opening angle is covered by at least $ 96 $ cells, an improvement of more than an order of magnitude compared to G22. Interestingly, as we show next, the magnetic dissipation remains, albeit it is continuously distributed throughout the jet rather than concentrated at the nozzle.
	
	\begin{figure}
		\centering
		\includegraphics[scale=0.19,trim=0 0 0 0]{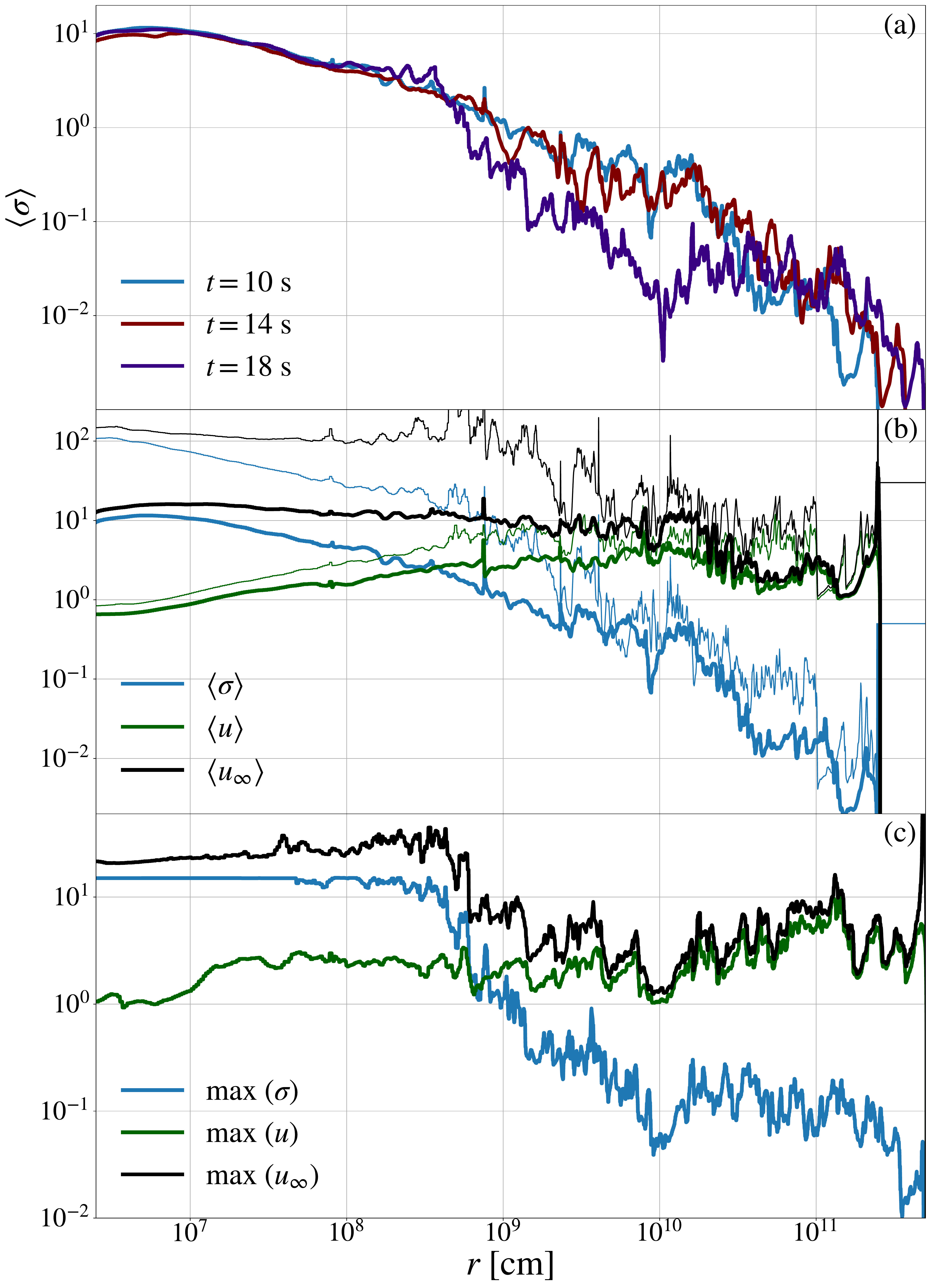}
		\caption[]{
		    \textbf{Panel (a):} Radial magnetization profiles of the jets with $ \sigma_0 = 15 $, calculated by taking the weighted-average magnetization over the radial energy flux excluding the rest-mass energy flux and fluid elements with $ u_\infty < 1 $. The profiles at different times indicate that the magnetization in the jets does not change over time and remains at $ \langle \sigma \rangle \sim 10^{-2} $ after breakout. \textbf{Panel (b):} Profiles of the weighted averages of $ \sigma $ (blue), proper velocity $ u $ (green), and asymptotic proper velocity $ u_\infty $ (black), shown for jets with $ \sigma_0 = 15 $ (thick lines) and $ \sigma_0 = 200 $ (thin lines) when the jet head is at $ r \approx 6R_\star $ ($t=10$ s). The jets are launched with high magnetization ($\sigma \gg 1 $), but the magnetic energy is efficiently converted into kinetic energy until $ r \approx 10^{9.5} $ cm, maintaining constant $ \langle u_\infty \rangle $. After this point, the mixing between the jets and the star increases, thereby reducing $ \langle \sigma \rangle $ further, and $ \langle u_\infty \rangle $ as well, as seen by their correlated values. The jets become mildly magnetized at $ r \approx 10^{10} $ cm. \textbf{Panel (c):} Profiles of the maximum values of the quantities in (b) for the jet with $ \sigma_0 = 15 $ at $ t = 18 $ s. It shows that some jet elements reach $ \Gamma \sim 10 $ after breakout with maximum $ \sigma \sim 0.3 $.
			}
		\label{fig:sigma}
	\end{figure}
	
	The bottom panel of Fig.~\ref{fig:early3d} shows a drop in the jet magnetization from $ \sigma_0 > 10 $ near the BH to $ \sigma \sim 1 $ within the inner $ 0.1R_\star $. Figure~\ref{fig:sigma}(a) shows the radial profile of the angle-average magnetization, $\langle\sigma\rangle$. To compute the angle average of a quantity, $\langle\dots\rangle$, we weigh its value by the radial energy flux, excluding the rest-mass energy flux and including only the relativistic outflow, i.e., the matter with asymptotic proper velocity $ u_\infty\equiv (\Gamma_\infty^2-1)^{1/2} > 1 $ , where $\Gamma_\infty$ is the asymptotic Lorentz factor. $\Gamma_\infty$ is the maximum Lorentz factor that would be attained if all of the gas thermal and magnetic energy were converted into the kinetic energy: $\Gamma_\infty \equiv -u_t\left(h+\sigma\right) $, where $ u_t $ is the covariant time component of the four-velocity, and $ h = \frac{4p}{\rho c^2}+1 $, where $ p $ and $ \rho $ are the comoving gas pressure and mass density, respectively. We show the magnetization profile at three times, taken after the jets break out of the star. The magnetization profile in the inner $10^9$ cm is rather smooth, with a similar shape at all times, whereas it shows a more complicated behavior at larger radii. At late times, the jet magnetization outside of the star saturates at values $10^{-2} \lesssim  \langle \sigma \rangle \lesssim 10^{-1} $.
	
	Figure~\ref{fig:sigma}(b) depicts the angle-average profiles of $\sigma, u$ and $u_\infty$ in the jet with the initial magnetization of $ \sigma_0 = 15 $ (thick lines) at $t=5$ s. At $r\lesssim 10^9$ cm, the proper velocity continuously increases while $u_{\infty}$ remains constant. This indicates that the magnetic energy is efficiently converted to kinetic energy of the bulk and accelerates the plasma. At $ r \gtrsim 10^9 $ cm, $u_\infty$ drops and shows a more erratic profile. We attribute this behavior to the entrainment of stellar material from the cocoon into the jet. The mixing of the light jet material with heavy stellar material reduces both $ \langle \sigma \rangle $ and $ \langle u_\infty \rangle $. A comparison with Fig.~\ref{fig:sigma}(a) suggests that the mixing becomes more significant at late times and is likely the main cause of the low $ \langle \sigma \rangle $ values outside of the star. Outside the star, the profile of $ \langle \sigma \rangle $ flattens, implying that the mixing stops after the ejecta breaks out of the stellar surface. The sharp drop at the farthest radii signifies the jet head.
	
	To examine the dependence of the simulation outcome on the initial magnetization, we show the results of an identical simulation with $ \sigma_0 = 200 $ as thin lines in Fig.~\ref{fig:sigma}(b). A comparison to the simulation with $ \sigma_0 =15$ shows that at the acceleration zone ($r<10^9$ cm), the behavior is similar, but the higher value of $\sigma_0$ enables acceleration to higher velocities. In the mixing zone, $r\gtrsim10^9$~cm, the mixing appears to have a stronger effect on the highly magnetized jet, as both $u_\infty$ and $\sigma$ drop faster with increasing radius. This result could be influenced by the higher susceptibility to the artificial density that is added by the simulation when the jet plasma hits the minimum density floor value. Nevertheless, we see that the results are qualitatively similar to the lower-magnetization case. Outside of the star, the magnetization and terminal proper velocity saturate at values of $\langle \sigma \rangle \sim 10^{-1}$ and $\langle u_\infty \rangle \sim 10 $ with peak values of $\sim 0.3$ and $\sim 30$, respectively. This indicates that jets with higher $ \sigma_0 $ can accelerate to higher $ u_\infty $ and $ \sigma $. The decrease of $ \langle u_\infty \rangle $ to $ \sim 0.1\sigma_0 $ in both models hints that in order for the jet to reach $ u_\infty \gtrsim 100 $, as expected in GRBs, the initial magnetization should be $ \sigma_0 \gtrsim 10^3 $.
	
	Figure~\ref{fig:sigma}(c) depicts the maximum values of $ \sigma, u $ and $ u_\infty $ at each radius at the end of the simulation with $ \sigma_0 = 15 $. This shows that parts in the jet maintain $ u_\infty \sim 10 $ and $ \sigma \gtrsim 0.1 $ after the breakout. Importantly, $ \sim 0.5\% $ and $20\%$ of the outflow energy is carried by plasma with $ \sigma > 0.1 $ when $ \sigma_0 = 15$ and $\sigma_0 = 200$, respectively (see also Fig.~\ref{fig:E_dist}(c)). Therefore, if $ \sigma_0 \gtrsim 10^3 $ as indicated above, then we anticipate $ \langle \sigma \rangle \sim 1 $ at the emission zone. We conclude that the jet has a hybrid composition at the photosphere, which includes magnetic and thermal components, implying that both play a role in the prompt $ \gamma $-ray signal (see \S\ref{sec:emission_mechanism}).
	
	\subsection{Post-breakout structure}\label{sec:postbreakout}
	
		\begin{figure}
		\centering
		\includegraphics[scale=0.243,trim=0 0 0 0]{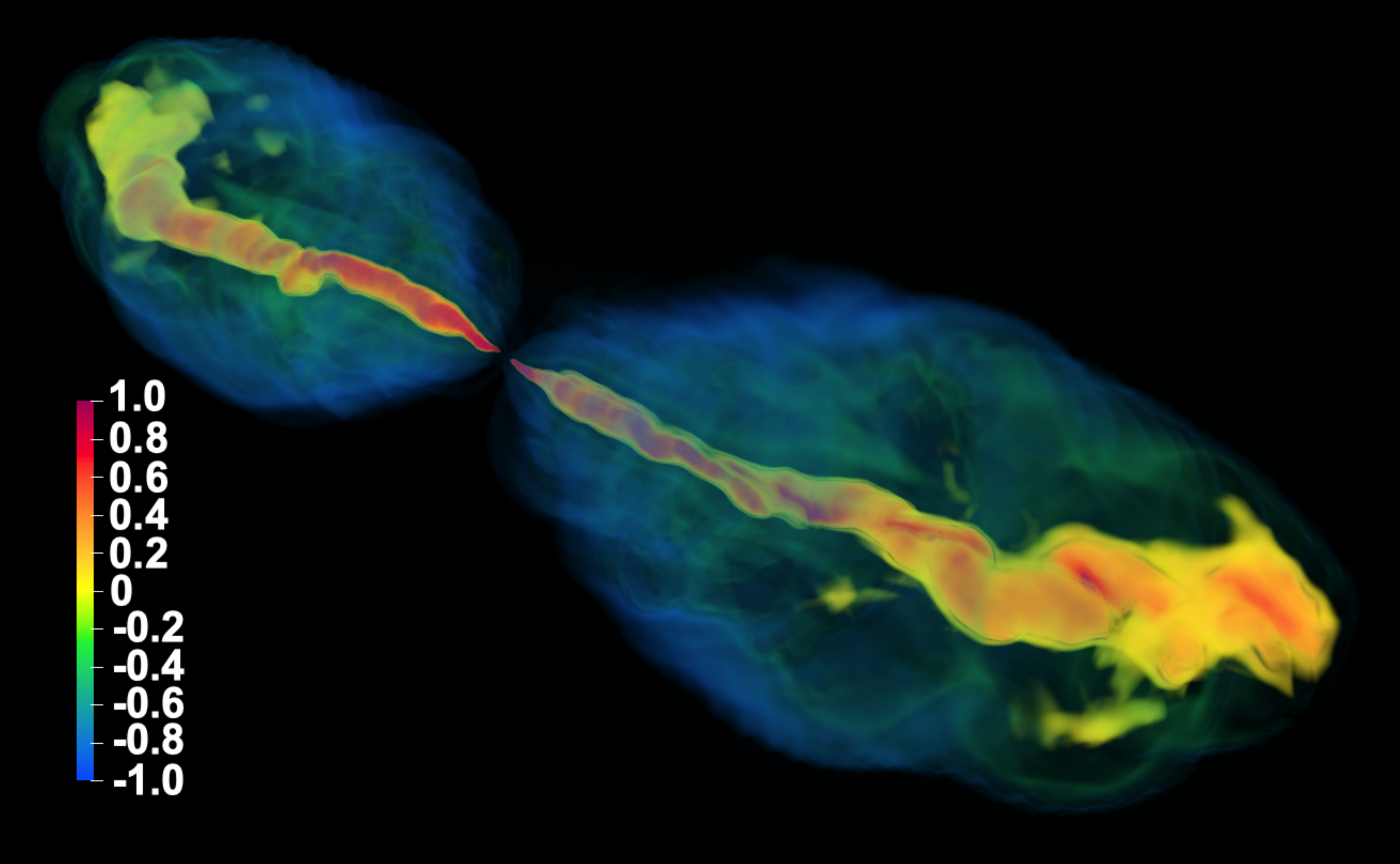}
		\includegraphics[scale=0.1875,trim=0 0 0 0]{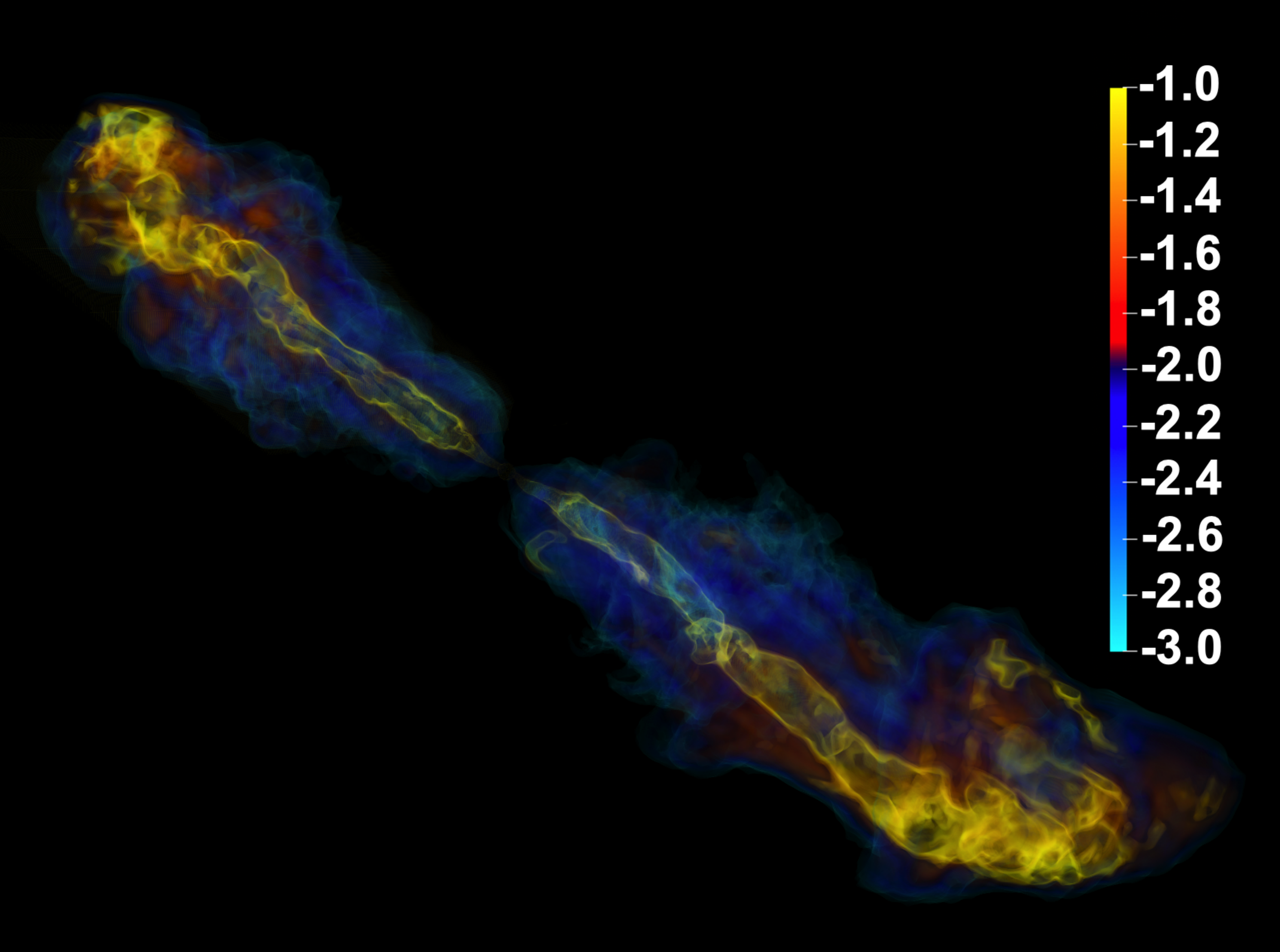}
		\includegraphics[scale=0.306,trim=0 0 0 0]{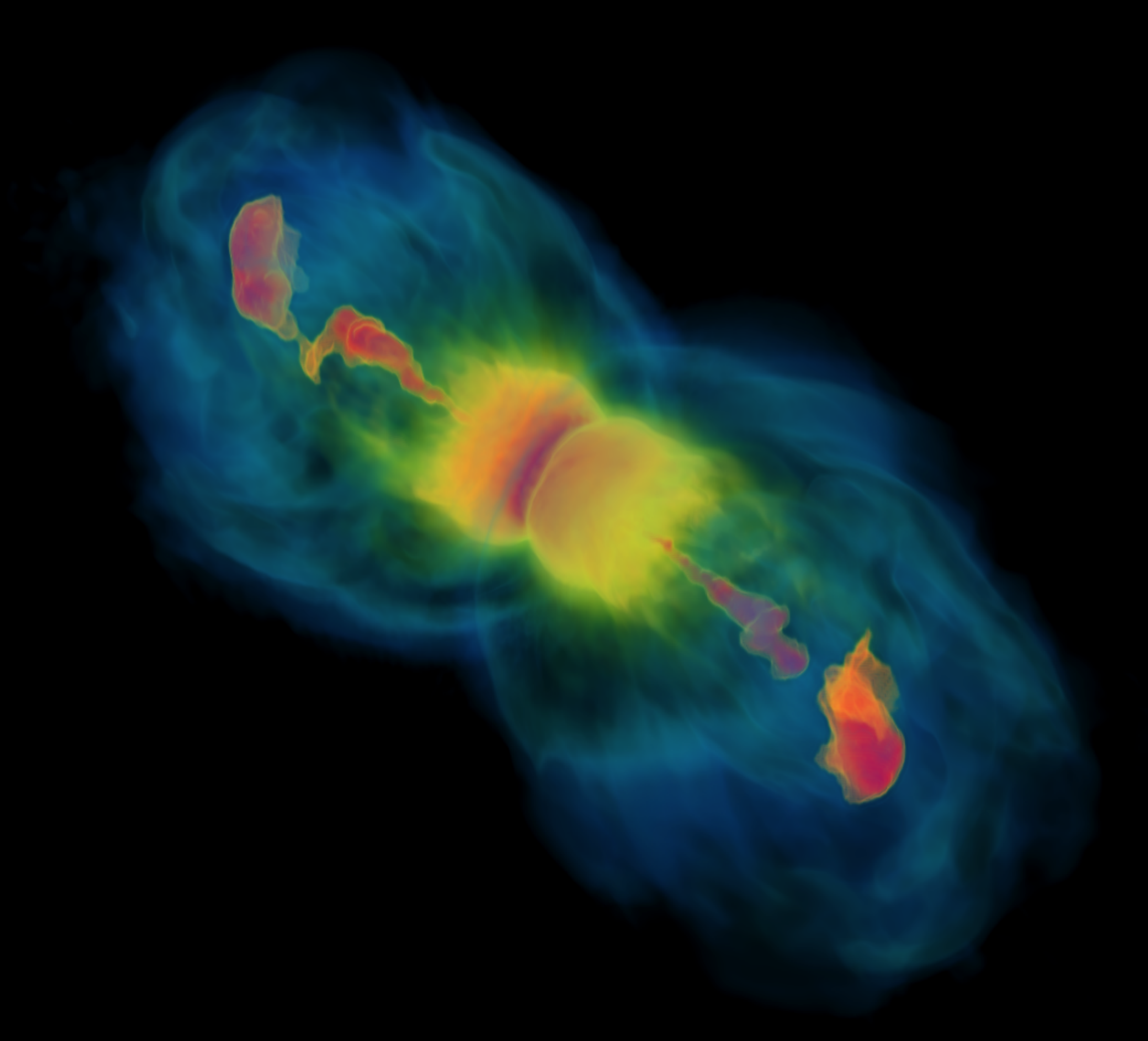}
		\caption[]{
			3D rendering in the simulation with $ \sigma_0 = 15 $ shows the logarithm of $ u_\infty $ (top) and $ \sigma $ (middle) in the jet (yellow)/cocoon (blue) structure when the jet head reaches $ r = 0.5R_\star $ at $ t \sim 1.5 $ s. It shows that at this radius the jet magnetization is already $ \sigma \lesssim 0.1 $. Bottom: The post-breakout outflow after 18 s, when the head of the intermittent jet (red) is at $ \sim 12 R_\star $ and the cocoon (green blue) explodes the star (yellow) entirely. The color coding is a combination of the mass density with $ u_\infty $ in order to show both the jet and the cocoon. Movies showing the full evolution of the outflow are available at \url{http://www.oregottlieb.com/collapsar.html}.
			}
		\label{fig:3djet}
	\end{figure}
	
	The propagation of the jet in the dense stellar envelope forms a hot cocoon that engulfs the jet. The top panel of Fig.~\ref{fig:3djet} shows a 3D rendering of the jet-cocoon structure, through the logarithm of the asymptotic proper velocity. While the jet maintains relativistic motion, the shocked jet material in the cocoon moves at mildly relativistic speeds. Some of the shocked stellar material in the outer cocoon is seen in subrelativistic velocities in dark blue. The middle panel depicts the logarithm of the magnetization, which decreases to $ \sigma \lesssim 0.1 $ in the jet, about halfway through its journey inside the star, while the cocoon maintains a weak magnetization of $ \sigma \sim 10^{-3} $. The bottom panel shows the final state of the simulation when the forward shock reaches $r \sim 12 R_\star $ $t \sim 18 $ s after the beginning of the simulation. Shown in green blue, the cocoon breaks out of the star (yellow) with a tilt angle of $ \sim 15^\circ $ with respect to the stellar angular momentum axis. The bipolar jets (red) are seen as wobbling blobs. The wobbling originates in the disk tilt, and the blobs emerge due to the intermittent nature of the central engine which increases the mixing \citep{Gottlieb2020a}. The blobs are moving at different velocities, and those that are moving in the same direction may produce internal shocks.
	
	The first 3D numerical characterizations of the post-breakout structure of GRB jets were calculated by \citet{Gottlieb2020b,Gottlieb2021a} for weakly magnetized and hydrodynamic GRB jets, respectively. Remarkably, they found that the angular profile of the isotropic equivalent energy can be modeled by a simple distribution, which consists of a flat core, followed by a power law in the jet-cocoon interface (JCI) and exponential decay in the cocoon. The main difference between hydrodynamic and weakly magnetized jets is the power law index, where weakly magnetized jets feature a steeper power law drop in the JCI, owing to suppressed mixing between jet and cocoon material due to the presence of magnetic fields.
	
	These works also considered the energy distribution per logarithmic scale of the asymptotic proper velocity, $dE/d\log{u_\infty}$, which at the homologous phase reflects the radial distribution of the outflow. They found that the relativistic part of the outflow has more energy if the jet is weakly magnetized, owing to the ability of magnetic fields to stabilize the jet against local hydrodynamic instabilities. The energy distribution in the cocoon ($ 10^{-2} \lesssim u_\infty \lesssim 3 $) was found to maintain a roughly uniform energy distribution in the proper velocity space and is independent of the jet magnetization. This result has been recently shown to have important observational implications for collapsars: (i) \citet{Barnes2018,Shankar2021} suggested that collapsar jets could be the powering mechanism for Type Ic supernovae (SNe Ic). However, because observations of those SNe indicate that the mildly relativistic component carries orders of magnitudes less energy than the subrelativistic ejecta, the flat distribution found in collapsar simulations cannot explain SN lightcurves, thereby ruling out jets as the sole source of SNe \citep{Eisenberg2022}. (ii) \citet{Gottlieb2022b} considered cocoon cooling emission as a possible source of the optical signal in fast blue optical transients (FBOTs). Because the radial energy distribution translates to the time evolution of the lightcurve, a quasi-uniform distribution suggests that the cooling emission in collapsars decays as $ L \propto t^{-2} $. Indeed, similar values are consistently found in all FBOTs with sufficient sensitivity for measuring the optical lightcurve to support this claim \citep[e.g.,][]{Margutti2019,Ho2020}. Nevertheless, the nature of FBOTs remains an open question and additional models have been shown to explain the nature of FBOTs using different assumptions \citep[e.g.][]{Margutti2019,Metzger2022}.

	\begin{figure}
		\centering
		\includegraphics[scale=0.24]{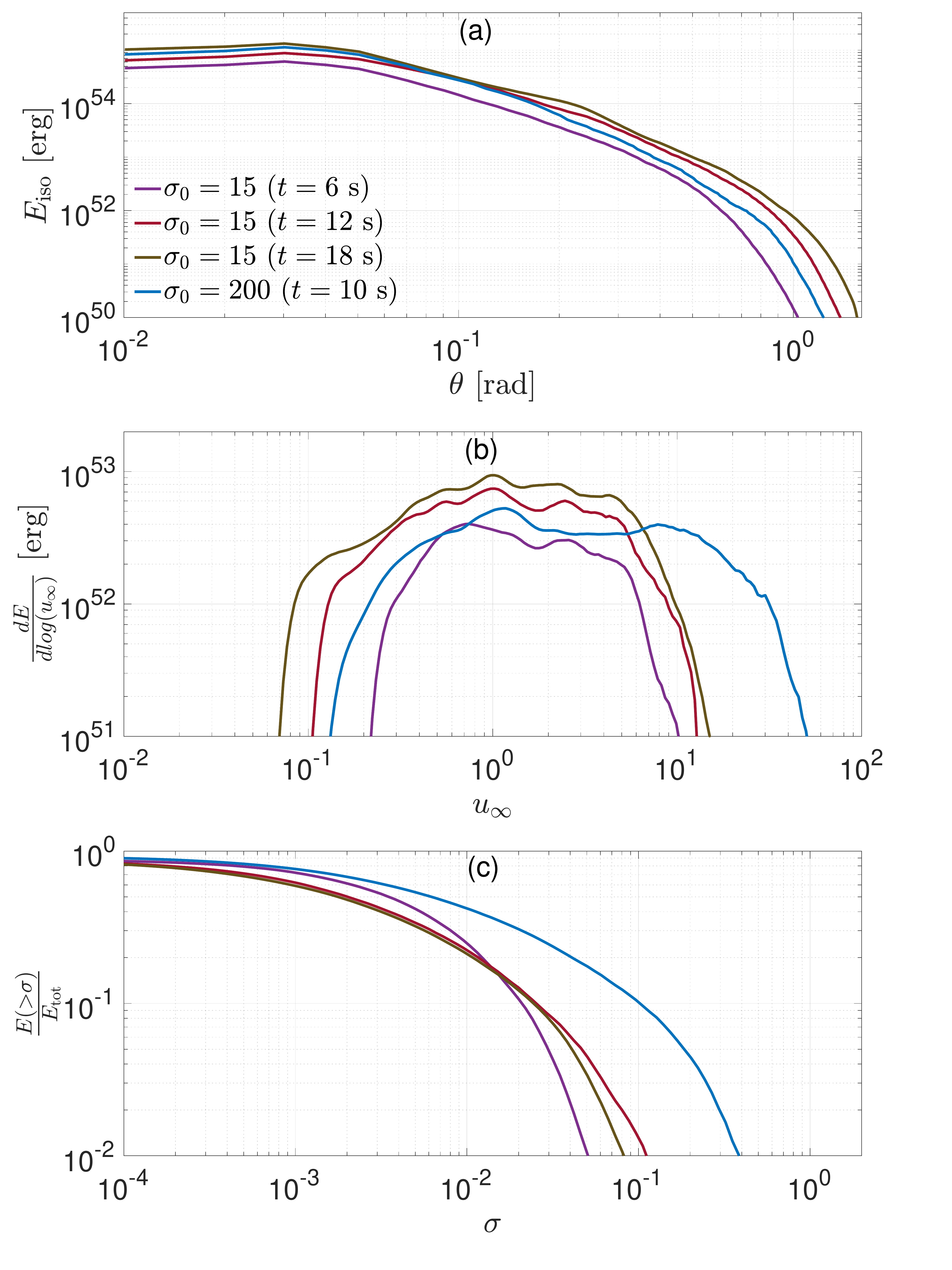}
		\caption[]{
		    The energy distribution of matter at $ r > R_\star $, excluding the rest mass energy, is shown according to various parameters. For the simulation with $ \sigma_0 = 15 $ we show the distribution at three different times: $ t = 6 $ s (purple), $ t = 12 $ s (maroon) and $ t = 18 $ s (gray). The blue line represents the distribution of energy in the jet with $\sigma_0=200$ at $ t = 10 $ s. \textbf{Panel (a):} The angular energy distribution integrated over the azimuthal angle of the outflow can be modeled by a flat core and a moderate power law followed by an exponential decay. \textbf{Panel (b):} The energy distribution per logarithmic asymptotic proper velocity is quasi-uniform irrespective to the choice of $ \sigma_0 $. Both angular and radial distributions are consistent with those of hydrodynamic jets \citep{Gottlieb2021a}. \textbf{Panel (c):} The fraction of energy in matter with magnetization larger than $\sigma$ out of the total energy. It can be seen that a significant amount ($ \sim 20\% $) of the outflow energy is carried by plasma with $ \sigma \gtrsim 10^{-2} $ ($ \sigma \gtrsim 10^{-1} $) when $ \sigma_0 = 15 $ ($ \sigma_0 = 200 $).
		}
		\label{fig:E_dist}
	\end{figure}
	
	Figure~\ref{fig:E_dist} complements the aforementioned works by presenting the first energy distributions at $ r > R_\star $ of highly magnetized jets launched by the rotation of the central BH. The evolution of the angular distribution of the outflow isotropic equivalent energy in Fig.~\ref{fig:E_dist}(a) indicates that the structure has reached the asymptotic stage in the simulation with $ \sigma_0 = 15 $, such that its shape no longer changes. The overall shape of the structure resembles the structure found for hydrodynamic and weakly magnetized jets \citep{Gottlieb2020b,Gottlieb2021a}. That is, the jet has a flat core with a characteristic angle of $ \theta_j \approx 6^{\circ} $ at which $ E_{\rm iso} $ drops to half of its value on the polar axis. The cocoon is characterized by an exponential decay and the JCI maintains a rather moderate power law $ \lesssim 2 $, similar to what was found in hydrodynamic jets. This result may come as a surprise, since weakly magnetized jets feature a steeper power law than hydrodynamic jets, owing to their stabilization effects. The reason for the similarity between our magnetized jets and hydrodynamic ones may lie in the absence of a self-consistent jet launching mechanism in previous simulations. The controlled injection of a jet via an artificial boundary condition in other works, avoids the stochastic behaviour seen in the bottom panel of our Fig.~\ref{fig:3djet}, which is affected by jet tilt, magnetic dissipation, intermittency of jet launching, etc. These phenomena increase the mixing between the jet and the ambient gas, and result in a flatter structure that is more similar to the less stable hydrodynamic jets. 
	
	Fig.~\ref{fig:E_dist}(b) shows the (radial) distribution of the energy in $\log$ of $u_\infty$ for matter outside of the star. The distribution is in agreement with previous results of a flat energy distribution\footnote{The flat distribution begins at the $ u_\infty $ that corresponds to the slowest material the broke out from the star at a given time, as can be seen by the temporal evolution of the minimal $ u_\infty $ of the flat distribution.} in cocoons of jets with different magnetizations, implying that this is a universal outflow distribution which is insensitive to the underlying physics. Therefore, our results support the conclusions of \citet{Eisenberg2022} by showing that Poynting-flux driven GRB jets cannot solely account for SNe Ic. It also provides a robust prediction that the expected lightcurve of cocoon cooling emission is similar to that observed in FBOTs \citep{Gottlieb2022b}, irrespective to the jet magnetization.
	The choice of $ \sigma_0 = 15 $ limits the outflow velocities to $ u_\infty \lesssim 15 $, inconsistent with those inferred from GRBs. For comparison, we show the energy distribution of jet material with $ \sigma_0 = 200 $ outside the star, when the jet head is at $ r \approx 10^{11} $ cm (blue line). The distribution remains flat, but the cut-off in the velocity is at $ u_\infty \sim 40 $. We note that the flat energy distribution of the cocoon is only weakly affected by the choice of $ \sigma_0 $ such that it is robust. We conclude that both angular and $\log(u_\infty)$ energy distributions are consistent with those of hydrodynamic jets, even to a better extent than with those of weakly magnetized jets.
	
	Finally, Fig.~\ref{fig:E_dist}(c) displays the cumulative energy distribution of matter that broke out of the star and has magnetization larger than $\sigma$, normalized by the total energy. Only a negligible amount of energy is carried by matter with $ \sigma \gtrsim 0.1 $ at $ t = 18 $ s in the jet with $ \sigma_0 = 15 $. However, in the model with $ \sigma_0 = 200 $ about $ \sim 20\% $ of energy carried by matter with $ \sigma \gtrsim 0.1 $. Such a significant amount of energy should have an important contribution to the prompt $ \gamma $-ray signal (the emission is discussed in \S\ref{sec:emission_mechanism}).
	
	\subsection{Jet tilt}\label{sec:tilt}
	
	\begin{figure}
		\centering
		\includegraphics[scale=0.225,trim=0 0 0 0]{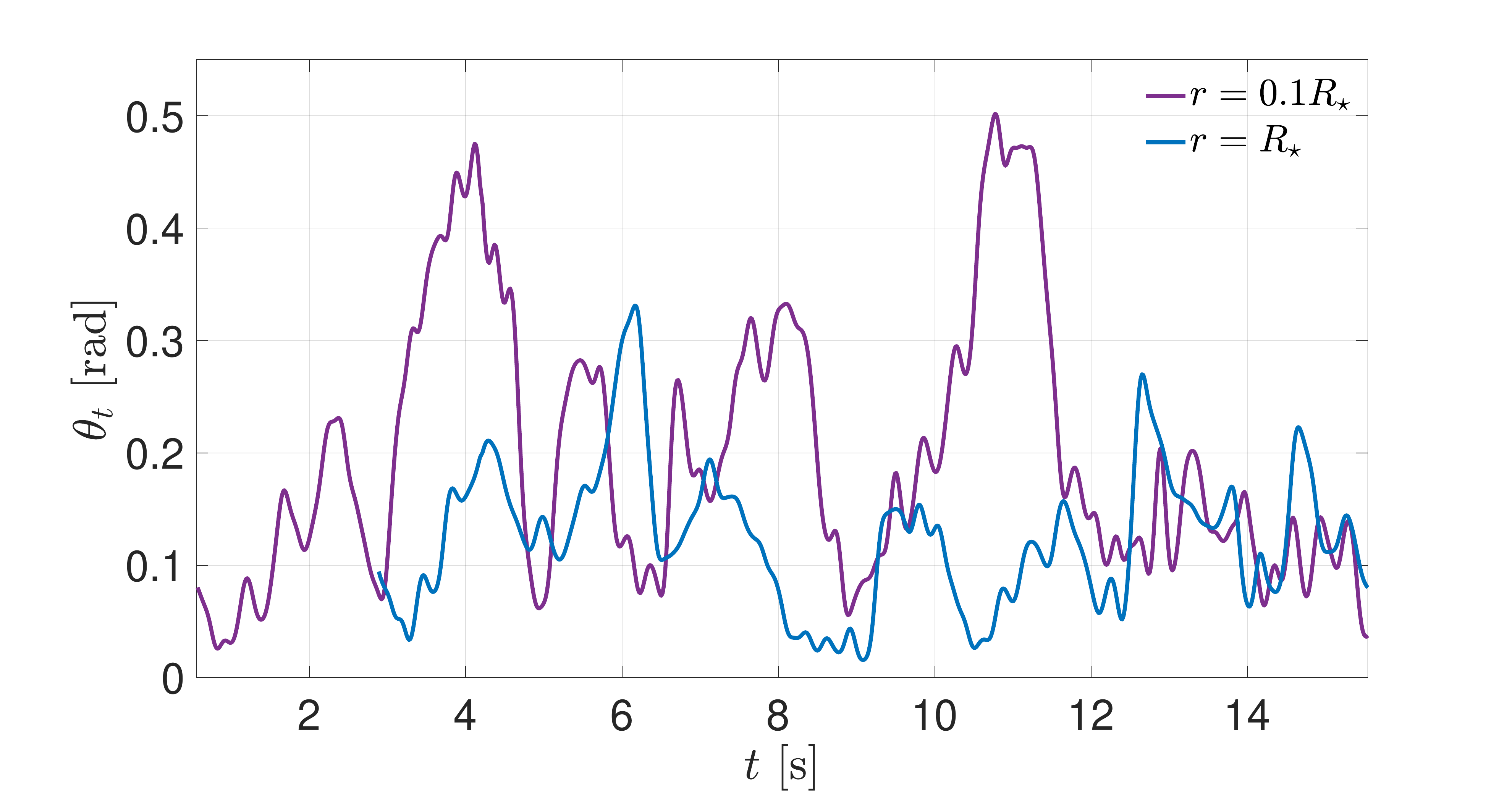}
		\includegraphics[scale=0.225,trim=0 0 0 0]{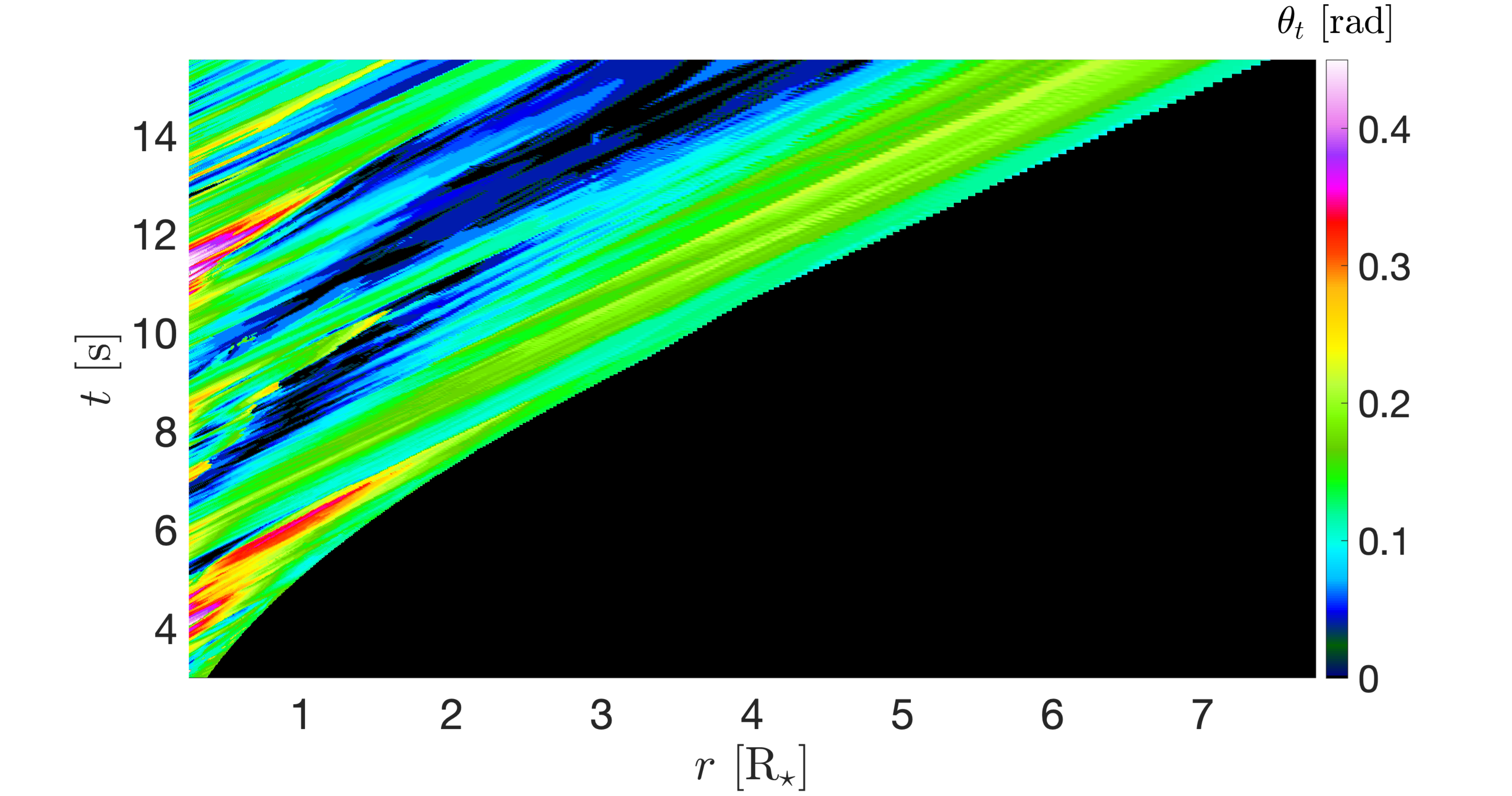}
		\caption[]{
			The time evolution of the jet tilt angle $ \theta_t $ in the simulation with $ \sigma_0 = 15 $. Top: The tilt angle deep inside the star (blue) and upon breakout (purple) shows that $ \theta_t $ is attenuated as the jet propagates in the star and its orientation is focused toward the stellar rotation axis. Bottom: Colormap of the jet tilt angle shows that it varies significantly inside the star, but is conserved after breakout along a light-like streamlines. The asymptotic jet tilt angle is $ \theta_t \lesssim 0.2 $ rad.
		}
		\label{fig:angle_evolution}
	\end{figure}
	
	In \S\ref{sec:prebreakout} we showed that the jet is launched with a considerable tilt with respect to the stellar rotation axis, but once it encounters the dense shoulders of the enveloping cocoon, it is deflected toward the low density region along the axis that was paved by the previous jet elements. Therefore, even though the jet initial orientation can be far from the $ \hat{z} $-axis, its final deviation from the axis is moderated by its interaction with the cocoon. To quantify the jet tilt angle evolution, we compute  as a function of radius the jet deviation from the polar axis, $ \theta_t(r) $, defined as the angle at which the maximum $ u_\infty(r) $ is measured, which we identify as the location of the jet axis.
	
	The top panel of Figure~\ref{fig:angle_evolution} shows the jet tilt angle $ \theta_t $ at $ r = 0.1R_\star $ (blue) and at $ r = R_\star $ (purple). At $ t \gtrsim 3 $ s the tilt angle at $ 0.1R_\star $ deviates significantly from the polar axis with $ \theta_t \lesssim 0.5 $ rad. The jet elements at $ r = 0.1R_\star $ reach the stellar surface after $ \sim 2 $ s, which corresponds to the shift in time between the blue and red peaks in $ \theta_t $. The collimation process that the jet orientation undergoes by the cocoon moderates the jet deviation from the stellar rotation axis such that the tilt angle amplitudes drop to $ \theta_t \sim 0.2 $ rad upon breakout from the star.
	
	The bottom panel of Fig.~\ref{fig:angle_evolution} shows a space-time diagram of $ \theta_t $ at each time and radius. The map confirms that $ \theta_t $ may decrease considerably at $ r \lesssim R_\star $, but outside the star the interaction between the jet and the dense cocoon is minimal, so that the jet maintains a roughly constant tilt angle along light-like streamlines (diagonal paths on the map).
	
	\section{Observational implications}\label{sec:emission}
	
	We have presented the results of the first 3D GRMHD simulation of collapsars that follow the relativistic jets from their formation at the BH to the photosphere (at $ \sim 10^{12} $ cm, see below). We discuss the significance of our findings in the context of observations; however a detailed study that computes the observational signature numerically is required and should be addressed in a future study.

	\subsection{GRB rate}
	
	GRB jets are typically considered to have an opening angle $ \theta_j $ around a fixed axis, with $ \theta_j $ that is significantly larger than the inverse of the jet Lorentz factor, $ \Gamma \gtrsim 100 $. After the jet breakout from the star, the blast wave interacts with the interstellar medium, its Lorentz factor decreases and eventually reaches $ \Gamma = \theta_j^{-1} $. At this point emission from the entire jet front reaches the observer and further jet deceleration leads to a jet break in the lightcurve. Adopting this picture, numerous studies \citep[e.g.,][]{Frail2001,Bloom2003,Guetta2005,Racusin2009,Fong2012} examined the jet opening angle, and obtained similar constraints of $ 0.07~{\rm rad} \lesssim \theta_j \lesssim 0.16~{\rm rad} $, based on which the local GRB event rate of $ \sim 1~{\rm Gpc^{-3}~yr^{-1}} $ \citep[e.g.,][]{Wanderman2010} translates to a total rate of $ \sim 100~{\rm Gpc^{-3}~yr^{-1}} $. The fraction of GRBs in SNe Ib/c also depends on the jet opening angle. It follows from the aforementioned estimates that GRBs likely exist in $ \sim 0.5\% - 3\%$ of SNe Ib/c. This result is also consistent with radio surveys, which provide an upper limit of $ \sim 10\% $ \citep{Soderberg2006}.

	While the above estimates of $ \theta_j $ are consistent with our simulations that show $ \theta_j \approx 0.1 $ rad, the aforementioned inferred GRB rates assume that the jet orientation does not change during the jet's lifetime. If jets in nature oscillate with $ \theta_t \sim 0.2 $, as found in our simulation, then their effective opening angle for detection is $ \sim \theta_t + \theta_j $ (Fig.~\ref{fig:emission}), making them $ \sim 10 $ times more frequent on the sky than previously estimated. This implies that we are able to observe $ \sim 10\% $ of all GRBs in the universe, and the total GRB rate drops by an order of magnitude to $ \sim 10~{\rm Gpc^{-3}~yr^{-1}} $, which is only $ \sim 0.1\% $ of all SNe Ib/c. One possible explanation for this scarcity is that most jets never manage to break out from the star to generate the GRB signal. This possibility was suggested before to explain the high rate of low-luminosity GRBs \citep[e.g.][]{1999ApJ...521L.117E,2001PhRvL..87q1102M,2003AIPC..686...74N} and the clustering of many lGRBs duration around $\sim10$ s \citep{2012ApJ...749..110B}. \citet{Margutti2014} and \citet{Nakar2015} proposed that jets fail to successfully drill through massive stars with extended envelopes (e.g. SNe Ib progenitors) and thus explosions of these stars are not coincidentally detected with GRBs. G22 showed that even among GRB progenitors that produce SNe Ic, only certain density, rotation, and magnetic profiles in the star support jet launching and breakout. The mildly relativistic outflow driven by these choked jets could power FBOTs \citep{Gottlieb2022b} whose estimated rate is $ \sim 10^3~{\rm Gpc^{-3}~yr^{-1}} $ \citep{Coppejans2020}.
	
	\subsection{Variability, quiescent times and periodicity}
	
	GRB lightcurves exhibit three characteristic timescales \citep{Nakar2002}. The first is the signal duration ($ \sim 10-100 $ s), which corresponds to the total work time of the engine. The second timescale is the rapid variability ($ \sim 10-100 $ ms), which corresponds to the stochastic fluctuations in the central engine accretion and launching mechanism and potentially to instabilities that develop during the interaction of the jet with the stellar envelope. Fig.~\ref{fig:properties}(b) shows the jet power that varies by about half an order of magnitude over the timescales of $ \sim 10{-}100$~ms, consistent with observations. Thus, we suggest that the rapid variability originates in the central engine, irrespective of the jet-star interaction (Fig.~\ref{fig:emission}), in agreement with \citet{Gottlieb2021b}.
	
	The third timescale corresponds to quiescent times \citep[$ \sim 1-100 $ s;][]{Ramirez-Ruiz2001,Nakar2002}, the origin of which is not fully understood. Our simulation shows that quiescent times naturally arise due to the jet tilt. When the jet is not pointing in the direction of the observer, its emission is beamed away, and the lightcurve becomes quiescent, potentially with some periodicity on timescales of $\sim 1-10$~s (e.g. GRB940210). The farther away the observer is from the polar axis, the smaller the ratio between the active signal time to quiescent time, and the lower the observed jet emission efficiency (Fig.~\ref{fig:emission}).	In addition to the quiescent times from the jet tilt, intrinsic temporal shutoffs of the engine can also produce quiescent times, albeit shorter. Figs.~\ref{fig:properties}(d) and \ref{fig:3djet} (bottom panel) illustrate $ \sim 1$~s fluctuations in the jet power, with relativistic jet blobs that are separated by $ \sim 1$ light-second of slow material. If the structure advances homologously to the photosphere, the slow material would produce a quiescent episode in the GRB lightcurve.
	
	\subsection{Emission mechanism}\label{sec:emission_mechanism}
	
    The origin of the prompt GRB emission is a matter of active debate, with many fundamental questions that remain unanswered to date, among which (i) is the observed emission subphotospheric or originating in optically thin regions? (ii) What is the underlying mechanism responsible for energizing the electrons that shape the non-thermal spectral tail? The two leading candidates for explaining the latter are particle acceleration in shocks\footnote{The acceleration can occur in collisionless shocks if the emission originates in optically thin regions, or in radiation mediated shocks if the emission is subphotospheric \citep[e.g.][]{2004ApJ...614..827R,2010ApJ...709L.172R,2011ApJ...733...85B}} \citep[see reviews by e.g.][]{1999PhR...314..575P,pir05,2006RPPh...69.2259M,2015PhR...561....1K} and acceleration in current sheets formed by magnetic reconnection
    \citep[e.g.][]{Thompson1994,Spruit2001,2003astro.ph.12347L,2008A&A...480..305G,Zhang2011,2016MNRAS.455L...6B}. These two mechanisms are directly connected to the energy composition in the jet, as the former requires strong shocks that can form in hydrodynamic or mildly magnetized flows ($ \sigma \lesssim 0.1 $), whereas the latter requires magnetically dominated flows with $ \sigma \gtrsim 1 $ \citep[e.g.][]{Sironi2015}. Thus, constraining the jet magnetization at the emission zone is an important step toward solving the prompt emission puzzle.

    Early models of steady, Poynting-flux-dominated jets showed that efficient conversion of magnetic energy to bulk motion is possible in narrow jets that are confined by an external medium, such that the jet magnetization can drop to $ \sigma \sim 1 $  \citep{tch08,kom09,Lyubarsky2009,Lyubarsky2010,Lyubarsky2011}. Further acceleration accompanied by a drop in $\sigma$ is possible if the jets undergo a sudden sideways expansion \citep{tch10b,kom10}, or if the jet is composed of individual pulses that expand in the radial direction \citep{Granot2011}. Such models require the jet plasma to maintain force-free conditions and avoid dissipation of energy via other processes, e.g. magnetic reconnection. Jet collimation often leads to the formation of narrow nozzles in which magnetic fields can become kink unstable and dissipate magnetic energy into heat \citep[e.g.][]{2012MNRAS.427.1497L,2009ApJ...700..684M,2012ApJ...757...16M,2019ApJ...884...39B}. In the case of continuous jets in dense media, \citep{Bromberg2016} showed that nozzle dissipation is a dominant process that can dissipate half of the jet magnetic energy into heat, rendering the outcome of the ideal acceleration models questionable. Further dissipation may continue above the collimation nozzle mediated by turbulence \citep{2019ApJ...884...39B}; however the state of the plasma that exits the star and when it reaches the emission zone remains an open question.
	
    Our simulations suggest that the jet intermittency and tilt increase the energy dissipation even further by allowing for the entertainment of heavy material into the jet, reducing the jet magnetization to values $ \sigma \sim 10^{-1} $ upon breakout from the star (Fig.~\ref{fig:sigma}), and possibly somewhat higher magnetization if $ \sigma_0 \sim 10^3 $ (see \S\ref{sec:dissipation}). By assuming the acceleration of a hydrodynamic-dominated jet, the optical depth evolves as $ \tau \propto t^{-3} $, we find that the photosphere is located at\footnote{Due to mixing, the photosphere is governed by the properties of the cocoon, rather than those of the jet. Thus, the value of $ 
    \sigma_0 $ is not expected to affect its location.} $ r \sim 10^{12} $ cm, similar to the photosphere found in hydrodynamic simulations \citep{Gottlieb2019}. If this magnetization persists in the jet through the propagation to the photosphere such that the jet keeps its hybrid composition, as indicated by Fig.~\ref{fig:sigma}, it renders both shock acceleration and magnetic reconnection as plausible energizing mechanisms. We note that the outflow is yet to reach a self-similar structure and the optical depth is likely to be somewhat different at later times; however our conclusion does not depend on these details.

	\begin{figure}
		\centering
		\includegraphics[scale=0.23,trim=0 0 0 0]{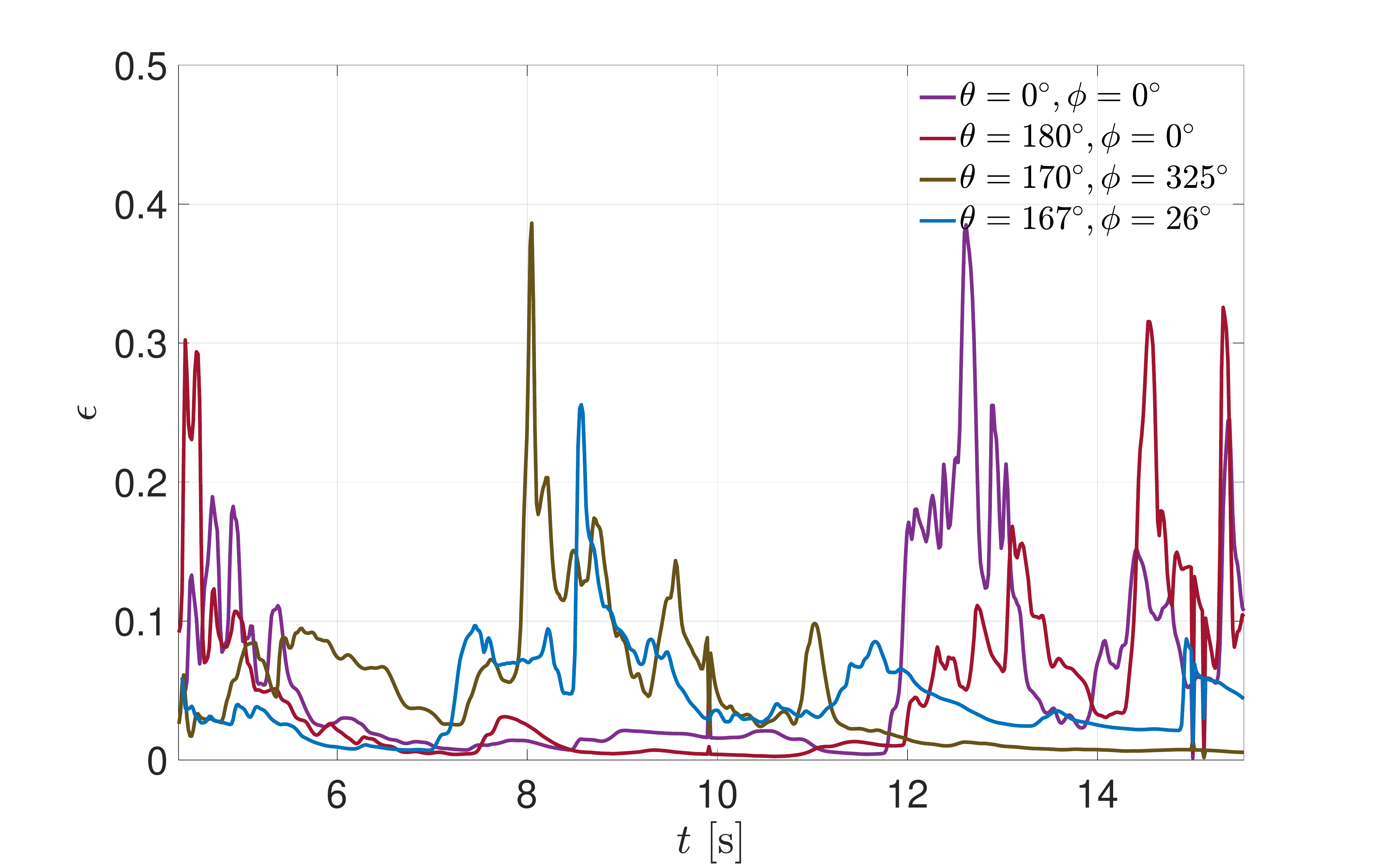}
		\caption[]{
            The radiative efficiency at $ r = 2R_\star $ for the jet with $ \sigma_0 = 15 $. The jet's wobbling motion is responsible for two features in the signal: i) long quiescent times at all viewing angles and ii) observers far from the polar axis, at $ \theta_t + \theta_j $ can detect high radiative efficiency. However, we find that in general, the efficiency decreases with angular distance from the axis.
		}
		\label{fig:emission}
	\end{figure}

	In \S\ref{sec:evolution} we showed that not only do the jets become partly hydrodynamic after breakout, but also their extended structure takes the same shape as that of hydrodynamic jets. It is therefore useful to examine the emission from these jets through previous numerical studies of 3D hydrodynamic jets. \citet{Gottlieb2019} found that owing to the collimation nozzle which converts the jet energy to thermal, jets re-accelerate at the larger radius of the collimation shock such that their coasting radius is above the photosphere. Consequently, hydrodynamic jets (and initially highly magnetized jets) inevitably produce a substantial photospheric component that can explain the high $ \gamma $-ray emission efficiency. Our simulations show that the above arguments also apply to highly magnetized jets. Specifically, we find that the post-breakout specific enthalpy allows $ 30\% - 80\% $ radiative efficiency (see Fig.~\ref{fig:emission}), depending on $ \sigma_0 $.

	One important difference that can arise between this work and that of hydrodynamic jets in \citet{Gottlieb2019} is at radii smaller than the dissipation radius, namely before the jet becomes hydrodynamic. In their paper, they found that the mildly relativistic collimation shock at the jet base produces a rest-frame temperature that is maintained at $ \sim 50 $ keV by pair production, which corresponds to the observed temperatures of a few hundreds keV. If at the collimation shock zone the jet still maintains $ \sigma \gg 1 $, as might be the case if $ \sigma_0 > 100 $, then the energy dissipation efficiency in the shock is low, and the resulting spectrum would be different.
	
	The hybrid magnetic and thermal composition with a magnetization of $ \sigma \sim 0.1 $ may broaden the thermal spectrum \citep[see also][]{Thompson2014} via reconnection and/or synchrotron emission. In addition to that, our simulation allows the formation of internal shocks that arise from the intermittent jet structure, which can similarly alter the spectrum if their radiative efficiency is high and they occur either above the photosphere or in a moderate optical depth \citep{Parsotan2018,Ito2019}. We conclude that our simulations suggest that the origin of GRB emission includes multiple components of photospheric, magnetic reconnection, and internal shocks, which can generate the observed Band function.

    We summarize the GRB emission in Figure~\ref{fig:emission}, where we show the radiative efficiency $ \epsilon \equiv (h-1)/h $ as measured at\footnote{Our choice of plotting the efficiency at $ r = 2R_\star $ comes from the need to show the temporal evolution, which we do not have at the photosphere at $ r \sim 10^{12} $ cm. While the efficiency at the photosphere is expected to be somewhat lower, we note that a higher $ \sigma_0 $ value will yield higher radiative efficiency. For example, we find in our simulation with $ \sigma_0 = 200 $ that $ \epsilon \approx 0.8 $ at $ r = 2R_\star $.} $ r = 2R_\star $ in the simulation where $ \sigma_0 = 15 $. Owing to the jet wobbling, observers at $ \theta  = \theta_t + \theta_j $ can detect high-efficiency emission. The jet motion also leads to the appearance of long quiescent times in the signal, while the jet intermittency results in a short timescale variability. The efficiency is typically higher close to the polar axis, as more episodes of jet pointing at this direction are expected.
    
	\section{Conclusions}\label{sec:conclusions}
	
	Using a novel GPU-accelerated AMR-enabled GRMHD code \textsc{h-amr}, we performed a collapsar simulation that is: (i) the first to follow jets from their self-consistent launching near the BH all the way to the emission zone at $ \sim 10^{12} $ cm, over 6 orders of magnitude in space and time, (ii) the highest-resolution 3D GRB simulation to date, and (iii) the first collapsar simulation that includes highly magnetized jets that emerge from a star. It is thus a major step forward in understanding the evolution of magnetized jets in general, and in collapsars in particular.

	The central engine launches relativistic Poynting-flux-dominated jets intermittently over $ \sim 10-100 $ ms timescales, consistent with the observed GRB lightcurve rapid variability. The jet is powered by magnetically saturated, MAD accretion and maintains the energy efficiency of order unity. The extended jet-cocoon structure reaches the photosphere with Lorentz factor $\Gamma \sim 30 $ (for $ \sigma_0 = 200 $) and jet opening angle of $\theta_j \sim 6^\circ $. The powerful outflow ($ \sim 10^{53}\erg $) unbinds most of the stellar envelope ($ \gtrsim 90\% $), but cannot explain SNe associated with GRBs due to the flat radial energy profile in the logarithm of the proper velocity space. A further investigation of the outflows that includes neutrinos may alter the SN-GRB picture and will be considered in a future work. The large fraction of unbound mass also implies that within the collapsar framework, the BH mass does not significantly change after the initial core-collapse, and therefore, the final BH mass should be similar to the stellar core mass. 
 
     The two main findings of our simulation are as follows:
	
	1. {\bf Jet tilt}: Infalling gas from the bound parts of the cocoon provide non-axisymmetric kicks to the disk and tilt it. Subsequently, the jet axis tilts as well and wobbles over a range of $ \sim 0.8 $ rad. As the tilted jet encounters the dense outer cocoon, it is refocused toward the low-density funnel drilled by the previous jet activity. Closer to breakout, the jet no longer changes its orientation so its post-breakout tilt angle is $ \theta_t \approx 0.2 $ rad. The jet tilt has important implications for observations of GRBs:
	
	(i) The tilt translates into an effective opening angle for detection of $ \sim \theta_t + \theta_j \approx 0.3 $ rad. This suggests that about $ \sim 10\% $ of the GRBs in the universe can be observed from Earth, implying that the total (intrinsic) GRB rate is $ \sim 10~{\rm Gpc^{-3}~yr^{-1}} $, about an order of magnitude lower than previous estimates. This raises the possibility that most jets fail to break out from collapsars. This can happen when some progenitor stars keep their outer layers of He and/or H that choke the jets, while others may just lack the conditions (e.g. magnetic and density profiles) essential for jet launching and breakout. (ii) Quiescent times naturally arise due to jet wobbling, when the jet is pointing away from the observer. For an observer who is far from the polar axis, only a few lower-efficiency episodes will be apparent. Shorter ($ \sim 1 $ s) quiescent times can also emerge due to the intermittent nature of the central engine.
	
	It is also noteworthy to mention the effect of such wobbling jets on the afterglow lightcurve. For a given observer who is aligned with the axis of one jet, the afterglow lightcurve will take the shape of a regular on-axis jet with a jet break that corresponds to the observed jet opening angle. This is because the emission of other jets is beamed away until their Lorentz factor becomes mildly relativistic, where the precise values depend on the offset of the jet from the observer. It then follows that jets that propagate outside the observer's line of sight may become visible only months after the explosion, and if these jets are comparably strong to the on-axis jet, their contribution to the lightcurve might not be detectable. Jets whose opening angles overlap with each other may lack a clear signature of a jet break in the lightcurve. We leave a full numerical study of the afterglow lightcurve from such structure to a future study.
	
	2. {\bf Magnetic dissipation}: Whereas the jet is launched as Poynting-flux dominated, its magnetic energy is continuously dissipated during its propagation. At small radii, the jet accelerates by converting magnetic to kinetic energy efficiently, with negligible amount of mixing between the jet and the star. As the jet propagates farther in the star, the mixing increases such that the jet magnetic (and kinetic) energy drops at a faster rate. The jet escapes from the star mildly magnetized, with $ \sigma \sim 10^{-1} $; after the jet breakout, the mixing weakens and the magnetization level remains steady. Thus, the jet structure is a hybrid composition of mildly magnetized and thermal parts, whereas its extended (radial and angular) structure is remarkably consistent with those found in hydrodynamic jets. In a companion paper, \citet{Gottlieb2022c}, we show that the picture is different for short GRBs jets that propagate in light ejecta. Those jets are subject to weak mixing and thus can retain $ \sigma \gtrsim 1 $ at the photosphere. The consequences of the jet becoming mildly magnetized for the observed emission are the following:
	
	(i) A substantial fraction of the magnetic energy is deposited as thermal energy in the jet, thereby enabling the jet to reach the photosphere with enough thermal energy to efficiently generate $ \gamma $-rays via photospheric emission. (ii) About $ 20\% $ of the post-breakout plasma maintains $ \sigma > 0.1 $ (when $ \sigma_0 = 200 $), which, together with internal shocks, may transform the spectrum from thermal to the observed Band function. (iii) The need to generate jets with $ u_\infty $ implies that $ \sigma_0 \sim 10^3 $, in which case we anticipate $ \sigma \sim 1 $ at the photosphere, such that magnetic reconnection will have an even larger contribution to the prompt signal.
	
	Finally, we outline a few limitations of our numerical setup that might affect our results. First, the simulations do not include any cooling scheme, e.g. via neutrino emission, which may change the disk evolution and alter the jet launching efficiency and duration. Second, our simulation does not model the phase between the core-collapse and formation of the BH. During this stage, a proto-magnetar is anticipated to form and launch magnetized outflows along the axis of rotation. Such outflows may alter the progenitor structure, primarily along the rotation axis, and mitigate the later relativistic jet propagation. Similar low-density regions along the axis may also form by neutrino-antineutrino annihilation or by the fast rotation of the progenitor star. Third, in our simulations, the post-breakout jet's isotropic equivalent luminosity is $ L_{\rm iso} \sim 10^{54}~{\rm erg~s^{-1}} $, which translates to prompt $ \gamma $-ray emission energy $ L_{\rm iso,\gamma} \sim 10^{53.5}~{\rm erg~s^{-1}} $, about an order of magnitude brighter than the peak luminosity of the observed lGRB distribution function \citep[see e.g.,][]{Wanderman2010}, and corresponds to the brightest known events. In a future study, we will explore the aforementioned caveats by modeling the progenitor structure self-consistently from the time of core-collapse to the BH collapse, including neutrino scheme M1 and examining a variety of jets. We stress that while those effects are important, our main conclusions are anticipated to remain similar, at least qualitatively.

	\begin{acknowledgements}

	We thank Fran\c cois Foucart, Daniel Kasen, Brian Metzger, and Ehud Nakar for helpful comments.
    OG is supported by a CIERA Postdoctoral Fellowship.
    OG and AT acknowledge support by Fermi Cycle 14 Guest Investigator program 80NSSC22K0031.
    AT was supported by NSF grants
    AST-2107839, 
    AST-1815304, 
    AST-1911080, 
    AST-2031997, 
    and NASA grant 80NSSC18K0565.
    AT and OB were partly supported by an NSF-BSF grant 2020747. 
    OB acknowledges support by an ISF grant 1657/18.
    DG acknowledges support from the Fermi Cycle 14 Guest Investigator Program 80NSSC21K1951, 80NSSC21K1938, and the NSF AST-2107806 grants.
    An award of computer time was provided by the Innovative and Novel Computational Impact on Theory and Experiment (INCITE) program under award PHY129. This research used resources of the Oak Ridge Leadership Computing Facility, which is a DOE Office of Science User Facility supported under contract DE-AC05- 00OR22725.
    The authors acknowledge the Texas Advanced Computing Center (TACC) at The University of Texas at Austin for providing HPC and visualization resources that have contributed to the research results reported within this paper via the LRAC allocation AST20011 (\url{http://www.tacc.utexas.edu}).
    This research was also enabled in part by support provided by Compute Canada allocation xsp-772 (\url{http://www.computecanada.ca}).

	\end{acknowledgements}
	
	\section*{Data Availability}
	
	The data underlying this article will be shared upon reasonable request to the corresponding author.
	
	\bibliography{refs} 
	
\end{document}